# When Molecular Dimerization Induces Magnetic Bi-Stability at the Metal-Organic Interface

*Rouzhaji Tuerhong[†], Franck Ngassam, Mebarek Alouani, Jean-Pierre Bucher[*]*

Université de Strasbourg, IPCMS UMR 7504, 67034 Strasbourg, France

**Abstract**

Two-dimensional (2D) metal–organic frameworks have been recently proposed as a flexible platform for realizing new functional materials including quantum phases. Here, we present a method to create metal-organic dimer complexes by on-surface assembly on a metal substrate using low-temperature scanning tunneling microscopy (STM) and spectroscopy (STS). We demonstrate that a dimer of Mn-Phthalocyanine $(MnPc)_2$ on a Ag(111) surface can be switched between two stable configurations upon a small conformational change controlled by STM manipulation. By means of density-functional theory calculations, it is found that the two conformations correspond to an antiferromagnetic (AFM) and a ferromagnetic (FM) state respectively. Directly coordinated Mn atoms of the dimer lead to an AFM-coupling whereas indirectly coordinated (shifted) Mn atoms lead to a FM-coupling. Rarely seen in a molecular-dimers with transition-metal atoms, the FM-AFM-FM transition is thus readily on-surface accessible. Furthermore, the two configurations of the switch are easily identified by their Kondo states, opening interesting routes in terms of both, writing (FM versus AFM states) and reading. These results pave the experimental route toward dimer-based materials with complex magnetic structures of potential interest for application in spintronics, logics and computing.

## 1. INTRODUCTION

Realizing some simple functions like a switch at the molecular scale has been a major challenge of molecular electronics and spintronics over years.[1] Studies have focused on layers of mononuclear complexes, mostly metallophthalocyanines that change their spin states upon adsorption of a gas such as NO,[2] CO,[3] $N_2$,[4] $H_2$.[5] The systems have been characterized by ensemble averaging technique (principally XMCD) and reversibility is generally attained by thermal desorption of the adsorbates. Other approaches are in a sense more straightforward since they directly address single molecules with a local electric field by means the tip of an STM, actually switching the molecules state.[6, 7] Whereas in the previous examples the functional purpose is achieved by only one molecule (spin center), it has been shown that it is sometimes of advantage to use more than one magnetic center to define one bit of information.[8] In particular, molecular dimers are entities that have their own electronic/magnetic properties that are (different from their constituents) of potential interest for application in spintronics, logics and computing.[9-11] However, to be used in devices, these objects need to be deposited and contacted to substrates, it is therefore important to know to which extent the properties of the dimers are altered or upon adsorption.

Whereas dimers made of bare metal atoms adsorbed on a metal surface have been studied by STM,[12, 13] similar studies on molecular dimers are scarce[14] despite the wide possibilities offered by coordination chemistry to tune molecular magnetic properties.[15-17] In this work we test the above concepts on manganese-phthalocyanine (MnPc) molecules that are made of a high-spin manganese atom in a ligand cage[18] and we examine the switching of a $(MnPc)_2$ dimer between stable conformations. For this purpose we investigate the different adsorption sites of a MnPc molecule on top of a first MnPc monolayer on Ag(111). Two stable sites were found by STM manipulation, corresponding to distinct configuration of the upper molecule with respect to the one below. The magnetic nature of the corresponding dimers is followed experimentally by STS. In particular the different magnetic states are examined by detecting the Kondo resonance induced by the screening of isolated spins by the conduction electrons of the substrate. We used DFT to compute the spin distribution of the stable dimers on Ag(111) and found that they correspond to ferromagnetic (FM) and antiferromagnetic (AFM) states respectively. It should be noted that the two configurations of the molecular switch are different from the ones found in the α- and β-bulk-phase,[19-22] see the Supporting Information (SI).

## 2. RESULTS AND DISCUSSION

### Dimer formation and conformational switching

To study the possible configurations of two molecules on a surface, we first deposit a full monolayer of MnPc monomers on a metal surface kept at $T = 300$ K. The MnPc molecules are known to form defect free monolayers on Ag(111) with a quadratic lattice whose parameters have been determined precisely.[23] Then the STM tip is used to manipulate isolated monomers that are adsorbed in the second layer, thus exploring accurately the equilibrium positions right above a first layer molecule to form a stable dimer. An example of molecular manipulation is shown in Figure 1a-c, where a single MnPc molecule is dragged out of the second layer island (a) and moved to a preferential adsorption site on the first monolayer (b). Details on the manipulation protocol can be found in the SI. The line profile of Figure 1c shows that the molecule has been moved to a place where it is shifted by a small amount along a well-defined direction with respect to the molecule located underneath thus forming one of the three stable configurations described in Figure 1d-f. These positions may also appear spontaneously after a molecular deposition slightly in excess of one monolayer and they can be tested back and forth by STM manipulation (see the SI).

The STM topographic analysis unveils three stable configurations. (i) In the on-top configuration (Figure 1d), the second layer molecule resides directly on top of the first layer molecule, so that its Mn atom is vertically aligned with the Mn atom of the lower molecule, thus forming a dimer in which the upper molecule has an azimuthal rotation around this vertical axis of 45° with respect to the lower one. (ii) In the shifted configuration (Figure 1e), the upper molecule is shifted with respect to the lower one along a Mn-N bond axis joining the Mn center to a pyrrole N atom (nearest neighbor N) by an amount of 175 ± 15 pm. The arms of the upper molecule remain aligned with the arms of the lower one (no rotation). (iii) In the bridge configuration of Figure 1f, two opposite Pc ligands of the second-layer MnPc molecule are located right above the Mn ions of two neighboring first-layer molecules and appear as two bright spots in the STM image, whereas the other two Pc ligands only show a weak contrast. As a result, the second layer Mn center is positioned half way between the two first layer Mn and appears as a small depression in the STM image. The frequency of naturally occurring configurations after dosing a small number of molecules on the first monolayer is distributed as follows: Shifted (30%), On-top (20%) and Bridge (50%). It is observed that the structures are stable below 400 K; a threshold is reached however during annealing at 500 K where the molecules of the second layer of the dimer desorb, leaving behind a pristine monolayer of MnPc. Among the three configurations observed in Figure 1d-f, the bridge involves three molecules and will not be discussed further here. The remaining two configurations correspond to $(MnPc)_2$ on Ag(111) made of two molecules either on-top or shifted with respect to each other. These two configurations are close in energy and will be discussed in detail.

We used density functional theory (DFT) calculations to compute the relaxed structures corresponding to the stable experimental configurations (see the Method section). Excellent agreement is found between the simulated equilibrium and experimental configurations. The optimized geometry of the on-top dimer shows that the Pc of the upper molecule is slightly bent and becomes convex, whereas the molecule in contact with the surface is planar (Figure 1g). The upper molecule of the shifted dimer on the contrary only shows a faint concave distortion (Figure 1h). Remarkably, the relative lateral shift value of 175 ± 15 pm for the upper MnPc molecule obtained from the STM experiment is in good agreement with the calculated in-plane molecular shift of 195 pm for the shifted dimer (see Figure 1h), which is equal to the calculated Mn-N bond length. This result indicates that the Mn center of the second-layer molecule is located right above a N atom of the first-layer molecule or vice versa, see the top view of Figure 1h. According to Figure 1, we will refer in the following to (1) and (2) for $1^{st}$ and $2^{nd}$ layer molecule respectively, that is lower and upper molecule with respect to the substrate. Therefore, the distance between lower and the upper Mn atoms along the z-direction is noted *d*(Mn1-Mn2). From GGA+U calculations, *d*(Mn1-Mn2) = 341 pm for the on-top configuration. For the shifted configurations of the dimer, a distinction must be made between *d*(Mn1-N2) = 307 pm and *d*(Mn2-N1) = 290 pm which are different, in the adsorbed state revealing a slight distortion of the dimer. The adsorption height of the dimer above the surface, that is the distance between the central Mn atom and the Ag atom beneath, is calculated to be 300 pm. A list of relevant geometric and magnetic parameters is given in Table S1 and S2. To highlight the influence of the Ag(111) substrate on geometrical and magnetic structure, each case has been investigated both in the vacuum and on the Ag surface. It can be seen that due to the interaction of the dimer with the Ag(111) surface, the MnPc1-MnPc2 distance in the dimer relaxes upon adsorption, from 335 pm to 341 pm for the on-top and from 278 pm to 290 pm for the shifted dimer.

The calculated energy for the dissociation of the $(MnPc)_2$ dimer into two separate MnPc molecules is 1.45 eV (139 kJ mol$^{-1}$) for the *on-top* case and 1.70 eV (164 kJ mol$^{-1}$) for the *shifted* case. The interaction is slightly larger for the shifted dimer due to the direct overlap of the two polycyclic molecules, contrary to the on-top dimer where this interaction is minimized due to the 45° rotation of the two molecules with respect to each other.

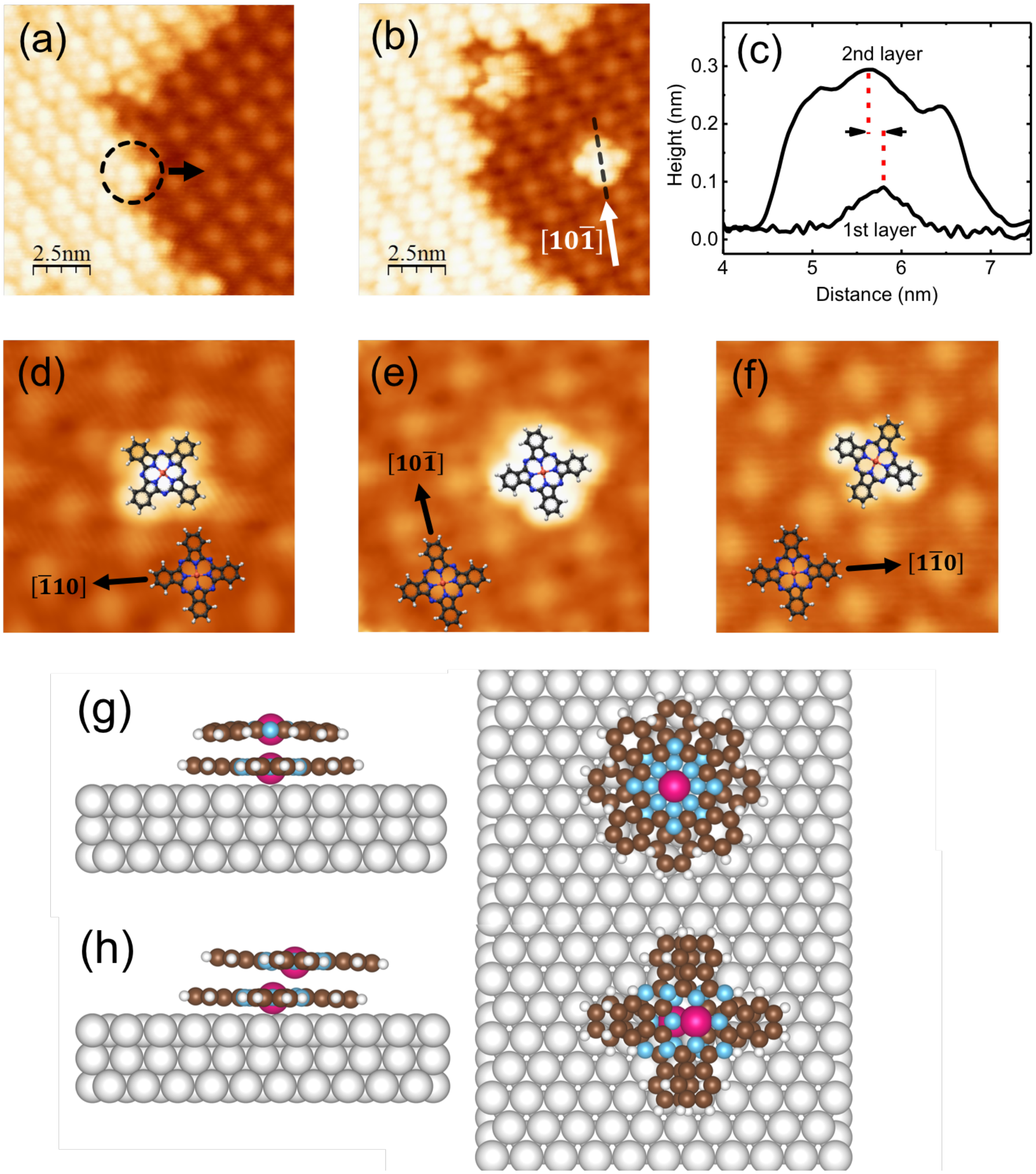

**Figure. 1** (a) STM images describing the molecular manipulation with the STM tip: A single MnPc molecule is dragged out of the second layer island and (b) moved to a preferential adsorption site on the first monolayer (imaging condition: –0.65 V, 10 pA). (c) Topographic line profile before and after manipulation, showing that the upper molecule is shifted upwards along the $[10\bar{1}]$ direction in (b) with respect to the molecule located underneath. The molecule shift is found to be 175 ± 15 pm. (d)-(f) STM images (6 nm × 6 nm) of single molecules above a MnPc monolayer on Ag(111) with (d) On-top, (e) Shifted and (f) Bridge configurations (–0.5 V, 10 pA). The bright STM contrast of the upper MnPc molecule is highlighted by the molecular model showing its configuration with respect to the underlying MnPc monolayer of which one molecule is highlighted by a model structure for clarity. The black arrows indicate the high symmetry directions of the underlying Ag(111) substrate. (g) *On-top* and (h) *shifted* structures of the MnPc dimer on Ag(111) obtained by DFT calculation. The color code for atoms is Mn: purple, N: Blue, C: Brown and H: White.

## Magnetic state and spin density distribution

To understand the effect of adsorption on the magnetism of the dimers (nature and strength), ab-initio DFT calculations were performed for both, the free standing and adsorbed dimers. The spin density for the equilibrium structures of the adsorbed phases are visualized in Figure 2, they clearly show that the antiferromagnetic (AFM) coupling between Mn is energetically favored for the on-top dimer (Figure 2a) whereas the coupling switches to ferromagnetic (FM) for the shifted dimer (Figure 2b). The nature of the coupling (FM or AFM), is unchanged upon adsorption as can be seen by comparison with the spin density for the vacuum-phase shown in Figure S4. An estimate of the exchange coupling $J$ between two MnPc molecules can be obtained by computing the energy difference $\Delta E_{\mathrm{spin}}$ between antiparallel and parallel spin state in either *on-top* or *shifted* configuration, and calculating $J = \Delta E_{\mathrm{spin}}/2$. The difference between vacuum and adsorbed phase is evidenced more clearly in Table 1, that shows a decrease in the magnitude of the exchange coupling $J$ when going from the vacuum to the adsorbed phase; proportionally more significant for the on-top dimer, from - 38 meV to - 11 meV, than for the shifted dimer, from + 122 meV to + 56 meV. Nevertheless, it is found that most characteristics of the free dimer are retained in the adsorbed phase, contrary to what has been calculated for a dimer on a more reactive magnetic substrate such as Co(001) for example.[24] This issue is most relevant for molecular spintronics where one usually wants to know if the properties of a molecular entity survive the adsorption on a substrate.[25, 26]

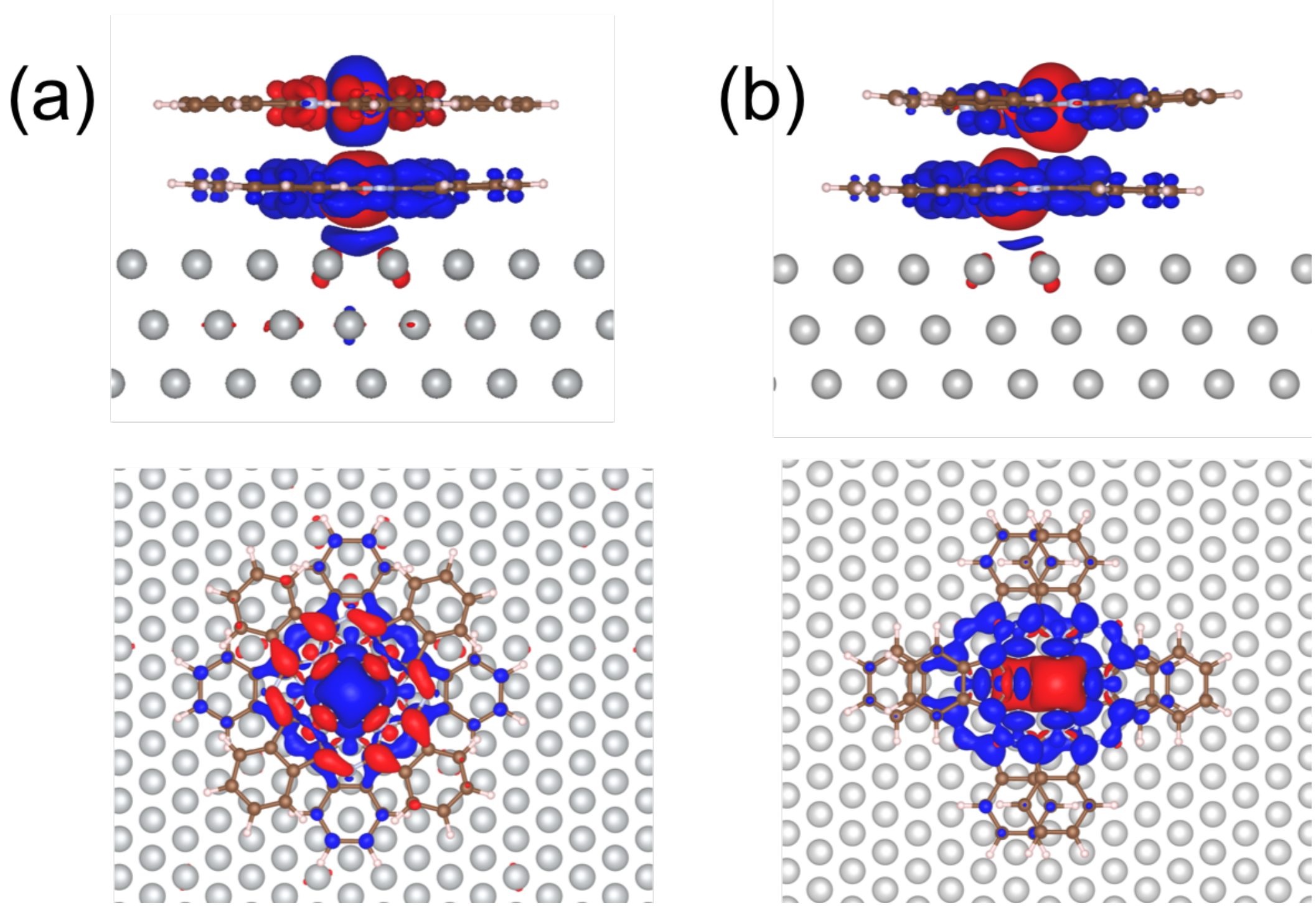

**Figure. 2** The spatial distribution of spin density in (a) *On-top* and (b) *Shifted* configuration of the MnPc dimer on Ag(111). The *on-top* configuration of the MnPc dimer couples antiferromagnetically (AFM) through direct exchange, and switches to the ferromagnetic (FM) coupling through superexchange in the *shifted* configuration. The red and blue colors are representing the spin-up and spin-down density, respectively. The isosurface value is 0.02 e Å$^{-3}$.

**Table 1**. Calculated magnetic coupling *J* given in meV for the two dimer configurations shown in Figure 2.

| | On-top | Shifted |
|---|---|---|
| In Vacuum | - *38* | + *122* |
| On Ag(111) | - *11* | + *56* |

**Wannier function analysis**

The magnetic coupling can be rationalized by analyzing the symmetries of the molecular orbitals involved in two dimer configurations. To visualize the orbitals involved in the magnetic exchange, we resort to the Wannier function analysis (see also the SI). In the *on-top* dimer of Figure 3a, the Mn of the lower and upper MnPc molecules are vertically aligned, and thus the exchange interaction between these Mn magnetic moments is dominated by a direct exchange coupling via the overlap of the Mn *3d* orbitals with favorable symmetries. The Wannier function projection onto the $3d$-orbitals shows that the main contribution comes from the $d_{z^2}$ orbitals (Figure 3b), protruding out-off the molecular plane. Hybridization between $d_{z^2}$ orbitals with a bonding character is the main contribution to the magnetic state of the Mn. Additionally, according to the $3d$-orbital occupations for the on-top dimer, it is legitimate to assume that it is the spin-½ of the singly occupied $d_{z^2}$ orbital (see Table S3) of each Mn that is forming a singlet state responsible for the AFM coupling between the two Mn in the on-top dimer. Other nearly singly occupied Mn $d_{xz/yz}$ and $d_{xy}$ orbitals are only marginally involved in the magnetic coupling due to the azimuthal angle of 45° between the two monomers which prevents any significant overlap between these orbitals (see also Figure S10a, top and side views). The strong $d_{z^2}$ orbital contribution is also confirmed in the bright contrast of the upper Mn atom of the on-top dimer in the STM image of Figure 4a (as compared to Figure 4b). Interestingly, the PDOS of Figure S2 shows that the $d_{z^2}$ state on Mn1 at $-$ 4 eV splits into multiple peaks upon adsorption, similar to the monomer in contact with the substrate.

In the *shifted* dimer configuration of Figure 3c, on the contrary, the Mn ions of the dimer are located right above or below a N atom (see Figure 1h). As a result, Figure 3d for the near Fermi PDOS of the shifted dimer on Ag(111) shows an important hybridization between N $2p_z$ orbitals and the Mn d-orbitals with a z-contribution (see also Figure S8a and b). This result is highlighted by the Wannier function analysis of Figure 3e, which confirms that this configuration of the two MnPc molecules enables an effective overlap between the Mn *3d* orbitals of the second-layer molecule and the N $2p_z$ orbitals of the first-layer, and vice versa. In addition, it is clear that an in-plane interaction takes place along the N-Mn bond pertaining to one and the same MnPc molecule of the dimer. Therefore, the shifted dimer suggests that the FM exchange interaction between Mn magnetic moments must be a 90˚ super-exchange interaction mediated by the N atom in the phthalocyanine rings. According to the orbital symmetry, both, the in-plane Mn $d_{xy}$ like orbitals and the out-of-plane Mn $d_{xz/yz}$, $d_{z^2}$ orbitals are involved in the FM coupling through a 90˚ super-exchange interaction, in agreement with the Goodenough-Kanamori rules for super-exchange.[27] This interaction is little perturbed by the adsorption as can be seen from the Mn1 and Mn2 PDOS that still look similar after contact (see Figure S7a), unlike the *on-top* dimer that keeps its vacuum PDOS for Mn2 and adopts the one of the monomer on Ag(111) for Mn1 (see Figure S5a and b). In addition, the calculation shows that the FM configuration of the shifted dimer has the smallest stray field i.e. induces the smallest polarization of the silver substrate (Figure S9) and is therefore less perturbed than the on-top dimer upon adsorption.

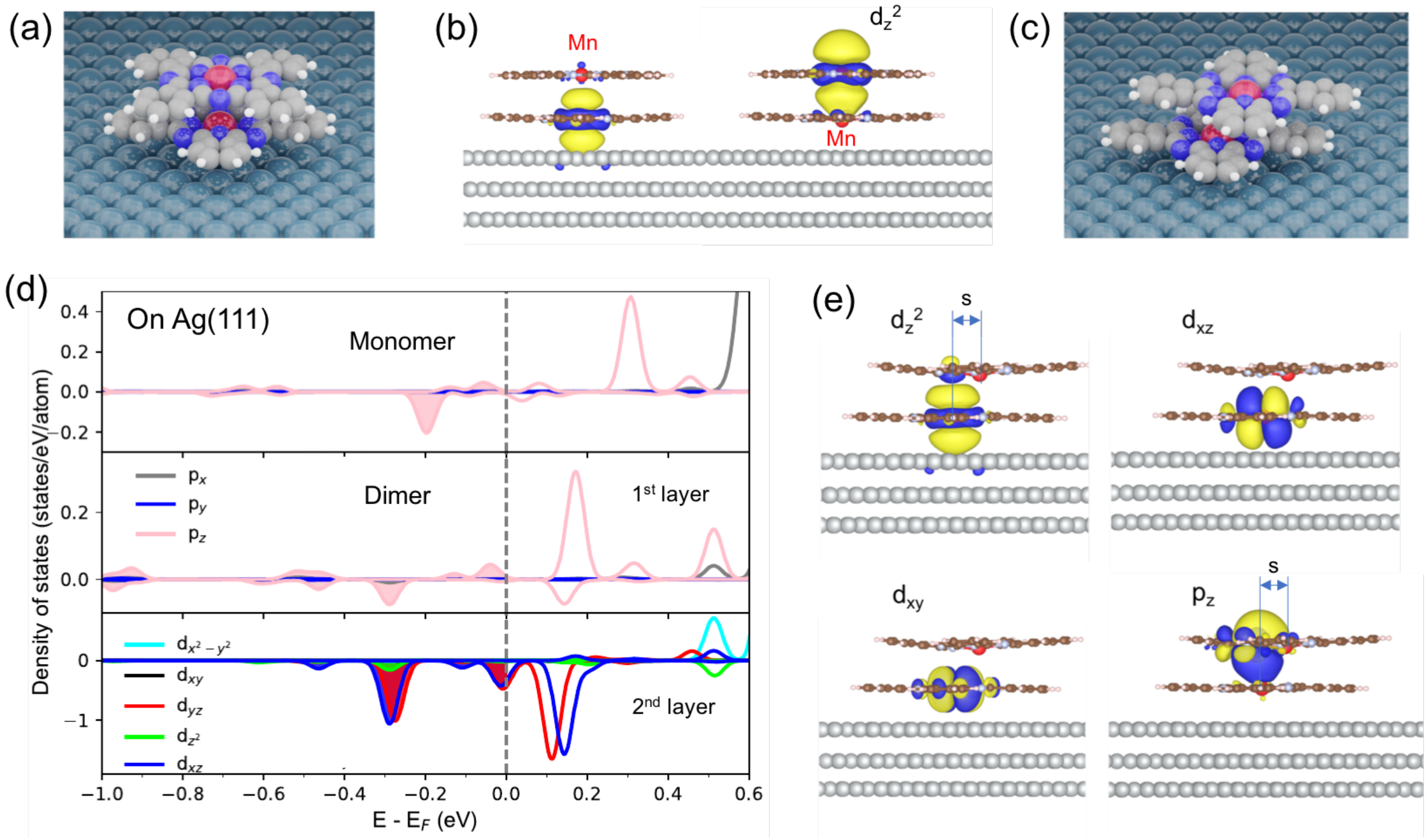

**Figure. 3** (a), (b) Adsorption geometry after relaxation and Wannier analysis for the on-top dimer: significant overlap is only possible for $d_{z^2}$ orbitals since other orbitals are less extended towards the z-direction, especially $d_{xy}$, $d_{x^2-y^2}$ but also for symmetry reasons in the case of $d_{xz/yz}$ orbitals since the 2nd layer orbitals of this kind are rotated by 45° with respect to the 1st layer thus strongly reducing the overlap. (c) Adsorption geometry after relaxation of the shifted dimer. (d) Shifted dimer PDOS: the d-orbital of the Mn in the 1st layer and p-orbital of the N-atom in the 2nd layer, just above the Mn, show significant p-d hybridization. (e) Wannier function analysis: the Mn $d_{z^2}$ orbital of the 1st layer is pointing towards the N-atom of the 2nd layer which in turn has its $p_z$ orbital pointing towards the Mn atom of the 1st layer allowing an effective overlap between the Mn 3*d* orbitals of the first-layer molecule and the N 2*p* orbitals of the second-layer, and vice versa. The color code for atoms is Mn: purple, N: Blue, C: Grey and H: White.

## Spectroscopic identification of the molecular switch

Experimentally, the STM spectroscopy provides additional many body information on the interaction of the MnPc dimer with the Ag substrate, not captured in the previous analysis. In particular, the Kondo screening[28] of the Mn 3d shell provides an additional means to distinguish the two states of the magnetic dimer. Figure 4c and d present typical differential conductance (*dI/dV*) spectra acquired in the near Fermi regime with the STM tip positioned above the Mn of the upper molecule (Figure 4a and b) for the on-top and the shifted MnPc dimer respectively. The spectra exhibit a zero-bias peak (ZBP) whose line shape and width strongly depend on the dimer configuration. Whereas a sharp peak with symmetric line shape at $E_F$ (zero bias) is observed for the on-top MnPc dimer (see Figure 4c), the peak broadens and becomes slightly asymmetric when the dimer is switched from the *on-top* to the *shifted* configuration (see Figure 4d). A good agreement is found between the temperature dependent *dI/dV* data and the Fermi-liquid analysis [29] as exemplified in the case of the upper MnPc of the shifted dimer (Figure 4e) and the MnPc in direct contact with the substrate (Figure 4f). This leads us to attribute the ZBP observed on this system to a Kondo screening of the local 3d magnetic moment by the conduction electrons from the substrate.

In the following, special attention is focused on the detailed analysis of the ZBP measured in the differential conductance, *dI/dV* spectra above the magnetic molecule. The shape of the ZBP is simulated by a Fano function: $dI/dV(V) = a + b\,(q + \varepsilon)^2/(1 + \varepsilon^2)$ with $\varepsilon = (\varepsilon V - E_0)/\Gamma_F$.[30] Here, $a$ is the background offset, $b$ is the amplitude of the Fano function, $q$ is the asymmetry factor of the Fano line, $E_0$ is the center of the Fano profile and reflects the energy location of the resonance. $\Gamma_F$ is the Fano line-width and reflects the characteristic half-width at half maximum (HWHM) of the Kondo resonance. The Fano line shape is a Lorentzian peak for $q = \infty$, and a Lorentzian dip for $q = 0$. In Figure 4c and d, the Fano fits (in red) are superimposed on the experimental spectra (black). The characteristic energy scale of the Kondo effect is normally expressed by means of the Kondo temperature $T_K$ that is related to the spectral width. In a real situation, however, one must take into account corrections for both, temperature[29, 31] and lockin voltage modulation (see the SI) so as to extract the intrisic values $\Gamma_K$ of the Kondo width and the corresponding Kondo temperature $T_K = \Gamma_K / k_B$. For the on-top dimer, the corresponding Kondo temperature and Fano-factor are $T_K = 60 \pm 4$ K and $q = 10$. This result must be compared with the shifted configuration of the dimer where $T_K = 125 \pm 7$ K and $q = -6.5$. The Kondo temperature of the dimer is therefore significantly increasing when going from the *on-top* to the *shifted* configuration.

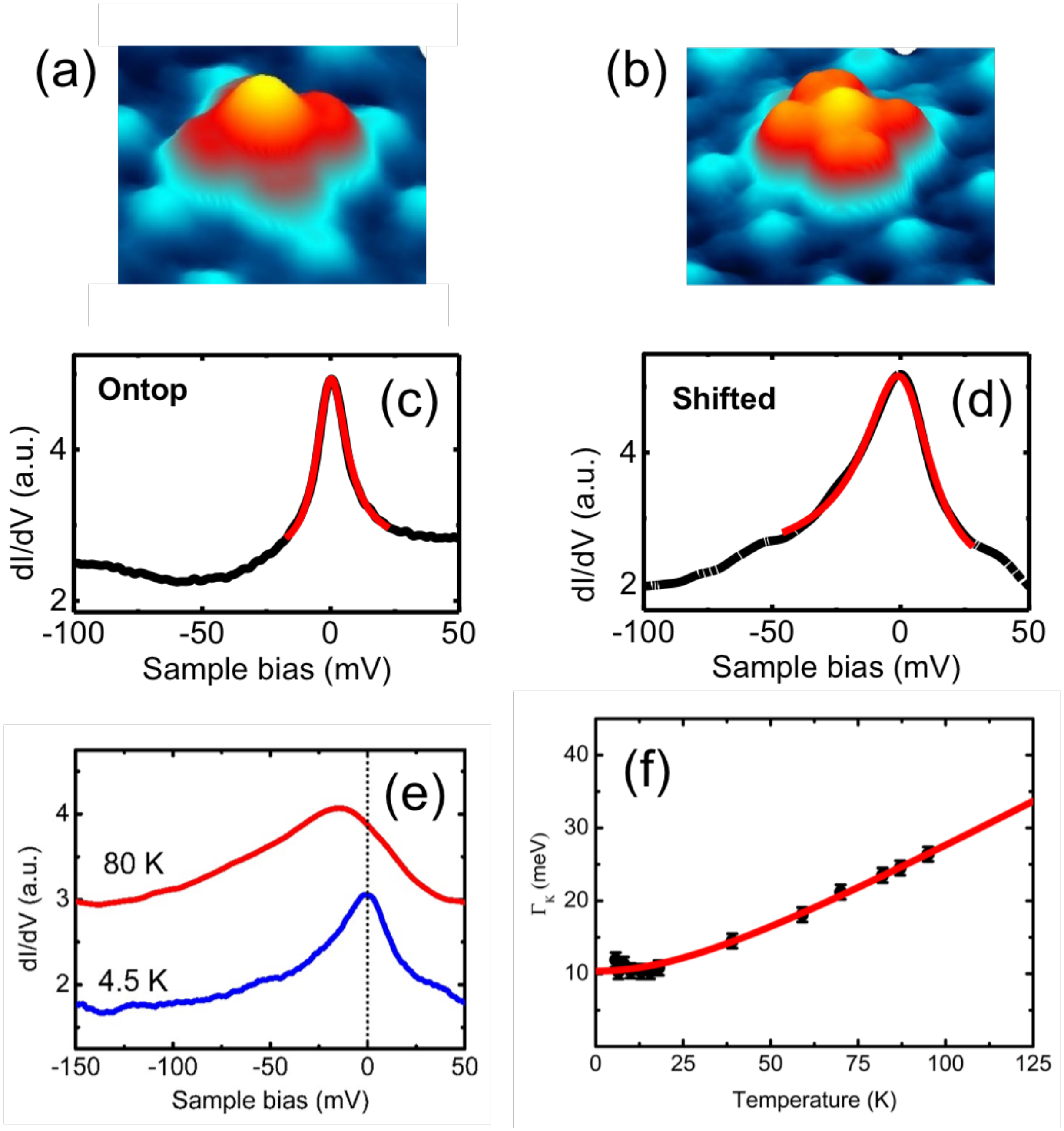

**Figure. 4** STM images (perspective view) and STS spectra for *on-top* and (a, c) and *shifted* (b, d) MnPc dimer on Ag(111). Near-Fermi *dI/dV* spectra (black) were measured above the Mn center of the second layer MnPc molecule. Each STS spectrum is fitted by a Fano function (red) (set point: –0.2 V, 100 pA). (e) Near-Fermi *dI/dV* spectra above the center of the upper MnPc of a *shifted* dimer at $T$ = 4.5 K and $T$ = 80 K (set point: –0.2 V, 100 pA). (f) Intrinsic width $\Gamma_K$ of the Kondo resonance as a function of temperature $T$ acquired above a Mn atom of a MnPc molecule in contact with the Ag(111). The black

dots are the experimental data. The solid red line shows a best fit to the function: $\sqrt{[(\alpha k_B T)^2 + (2k_B T_K)^2]/2}$, of the Fermi-liquid model[29, 31] see the SI. For $T >> T_K$, the width approaches a linear behavior with a slope $\alpha = 5.85 \pm 0.1$.

The line shape evolution between the two states of the dimer-switch is quantified by the difference in the Fano fitting parameters, thus underlining that the orbitals and the tunneling path involved in the Kondo physics are different. In particular, the lower $T_K$ and the large positive line shape parameter $q$ of the on-top dimer indicate a relatively weak coupling of the second-layer molecule to the metal substrate, and enhanced tunneling into the resonant state of the second-layer spin center. On the other hand, the ZBP of the shifted dimer in Figure 4d shows a Kondo temperature and a q value (including sign) that are very close to the ones observed for a single MnPc molecule residing directly on the Ag(111) substrate (see the SI). This speaks in favor of a very similar Kondo behavior for the first-layer and second-layer shifted molecules on Ag(111), which may look surprising owing to the fact that in both cases the MnPc molecules are subject to a different electronic environment. However, this experimental fact indicates that the Kondo screening of the MnPc spin moment of the upper molecule is almost as efficient as the one of the on-surface MnPc molecule in the first layer.

Whereas the Kondo resonance observed for the on-surface MnPc/Ag(111)[23] is attributed to the exchange interaction between the MnPc molecular spin and the itinerant conduction electrons of the Ag(111) substrate, this interpretation is unlikely to be fully valid for the Kondo peak of the second layer Mn of the dimer on Ag(111) and other mechanisms need to be analyzed: (i) The exchange coupling between the Mn1 and the Mn2 center may act as a relay to convey the spin-moment information to the substrate, so that it is the Mn1 spin that is ultimately screened by the conduction electrons of the substrate. (ii) The upper Mn2 spin is screened by the Ag substrate via an indirect interaction through the ligands and the bottom molecule (iii) The exchange coupling between the two Mn is so strong that the molecule can be considered as a single macro-spin that in turn is screened by the substrate.

The third option (iii) is unlikely since the total magnetic moment of the on-top AF dimer would then tend to zero leading to an absence of Kondo effect. To decide which option of (i) or (ii) applies, we need to compare the magnitude of the exchange coupling $J$ with the Kondo exchange interaction energy $J_K$ accounting for the coupling between the local magnetic moment of the upper Mn and the conduction electrons of the underlying MnPc/Ag(111) system. Our measurements of the Kondo temperature $T_K$, allows us to estimate $J_K$. Note that in both, *shifted* and *on-top* geometries, the first layer MnPc exhibit the same conformation with respect to the Ag(111) substrate. For sake of simplicity we consider that among the possible Kondo-channels of Table S3, only the $3d_{z^2}$ orbital with a single spin-1/2 occupancy is Kondo-screened.[32, 33] Therefore, the following relationship applies: $k_B T_K = De^{-1/2\rho J_K}$,[34] where $\rho$ and $D$ can be taken in first approximation as the Ag(111) density of states and conduction bandwidth. With $D \approx 5.5$ eV,[35] and assuming $\rho = 2/D$, we obtain: $J_K \approx 220$ meV and $J_K \approx 190$ meV respectively for the *shifted* and *on-top* case. It is found that the Kondo exchange interaction $J_K$ of the Mn with the conduction electrons is larger than the $J$ interaction between Mn1 and Mn2 obtained above (see Table 1), however, due to the margin of error of the above calculation, option (i) cannot be excluded completely. Option (ii) based on the indirect interaction of the upper Mn $3d$ spin moment with the substrate through the MnPc/Ag(111) composite-interface may be a better option, though. To verify this hypothesis, we compute the PDOS on the Ag atoms just underneath the first MnPc molecule of $(MnPc)_2$ on Ag(111). As shown in Figure 5, the bottom as well as the upper Mn d-orbitals are hybridized with the Ag(111) substrate. Note that for the shifted dimer, the features associated with occupied d-orbital states (negative energy) are seen in the same approximate energy range as the states of the two Mn. The partial occupancies of the *s*, *p* and *d* orbitals reveals furthermore that the interaction between the d-shell states, the ligands and the substrate is mediated through atomic hybrid orbitals with mixed *s*, *p* and *d* character. Similar chains of interactions between spin moments and substrates via the ligands have been invoked to explain the Kondo screening observed over double-decker molecules on Ag and Cu substrates.[36, 37]

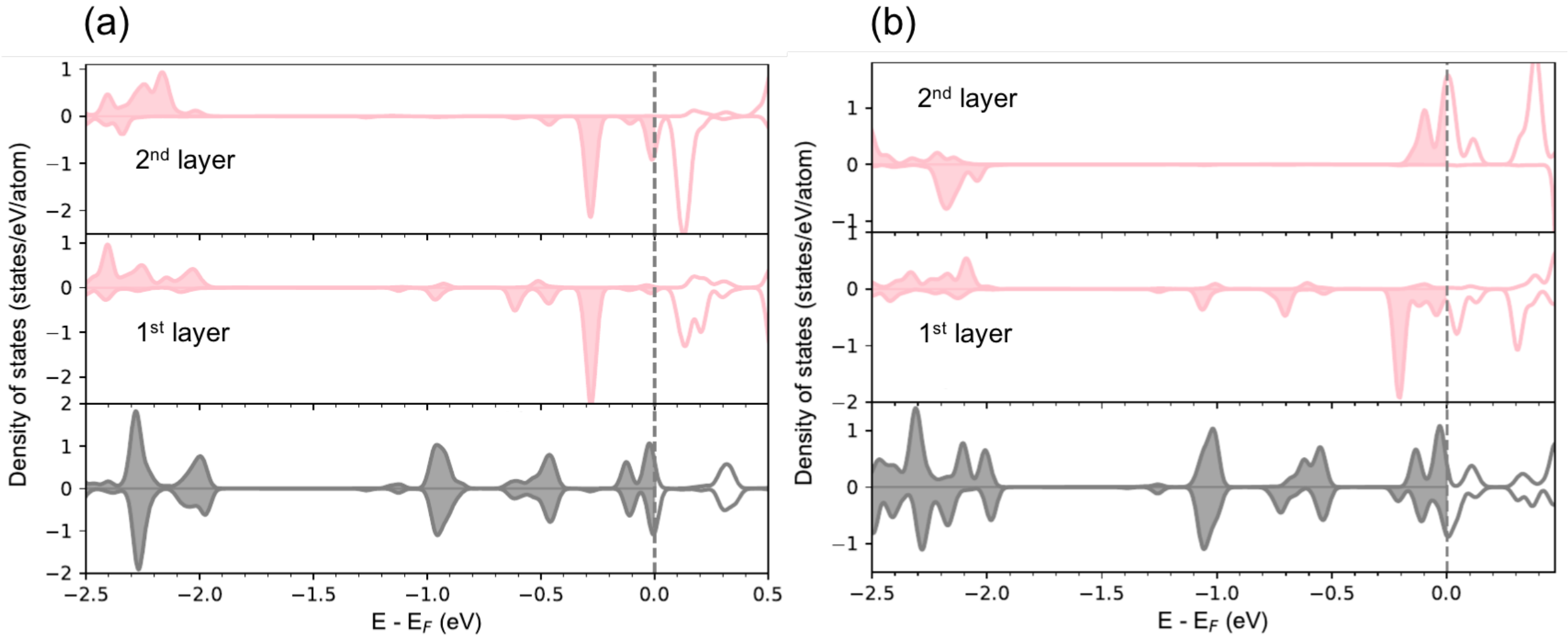

**Figure. 5** (a) Shifted and (b) On-top dimer. PDOS on the Ag atoms just underneath the first MnPc molecule of $(MnPc)_2$ on Ag(111) (lower panel) and PDOS on the Mn of the lower and upper molecule respectively (upper panels). The bottom as well as the upper Mn are hybridized with the Ag(111) substrate. Note that features associated with occupied d-orbital states (negative energy) are seen in the same approximate energy range as the states of the two Mn.

As a matter of fact, we find that for each MnPc molecule of the dimer, the ligand contributes a small (but not negligible) amount to the magnetization with -0.4 $\mu_B$ opposite to the contribution of the Mn with 3.5 $\mu_B$, see Table S1. The involvement of the Pc ligand in the Kondo screening has been clearly evidenced on single MnPc molecules in contact with the surface where the Kondo resonance extends over the Pc ligand.[23, 38] Accordingly, it is tempting to conclude that the Kondo coupling of the upper Mn spin moment to the π-electrons of the Pc ligands of the first-layer MnPc contributes as well to a spin-electron coupling mechanism, hence to the Kondo resonance of the upper MnPc molecule. This is in contrast to the molecular spin of CoPc on Pb(111) which is completely quenched due to the charge transfer to the Co center of the first CoPc layer which then acts as an insulating and spin quenched buffer layer for additional CoPc layers.[39, 40] The monolayer MnPc/Ag(111) instead is conducting and spin-active, this confers the MnPc bilayer on Ag(111) very interesting electronic and spintronic properties among the family of metal phthalocyanine multi-layers on metal surfaces.

## 3. METHODS

*Experimental methods.* Experiments were performed in a UHV system with a base pressure p < $10^{-10}$ mbar, equipped with a low temperature scanning tunneling microscope (Modified CreaTec LT-STM). The Ag(111) single crystal was cleaned by repeated cycles of $Ne^+$ ion bombardment followed by thermal annealing at 800 K. The MnPc molecules were sublimed from a home built ceramic cell. Before the deposition, the MnPc powder (purity > 95%) is degassed for several hours. Molecules have been deposited at a rate of 0.2 ML/min at an evaporation temperature of 250 °C on the silver substrate kept at room temperature. The sample is then transferred into the STM chamber. All STM/STS measurements have been performed at T = 4.5 K. Tungsten tips are flash annealed in UHV and conditioned by indentation into a Au substrate. STM images were acquired in a constant current mode and analyzed with the WS × M software. The bias voltage is applied to the sample with respect to the tip kept at a virtual ground. The *dI/dV(V)* spectra were acquired by a lock-in detection of the tunneling current, with a modulation voltage of 3 mV peak-to peak at a frequency of 900 Hz superimposed to the sample bias voltage. The *dI/dV* data are analyzed with the Origin 8.1 software.

*Theoretical Simulation Methods*. To compute the structural and magnetic properties of the $(MnPc)_2$/Ag(111) dimer system we performed density functional theory (DFT) calculations by means of the Vienna ab-initio simulation package[41] (VASP) and used the projector augmented wave (PAW) method[42] as implemented in VASP. The PAW method is an efficient method for the calculation of electronic structure of materials. The advantage of this method is that it exploits the simplicity of the pseudopotential method[43] while it also describes correctly the wave functions in the augmentation region. In the PAW method, the Bloch function is a linear combination of atomic functions in the augmentation region and smoother functions, plane waves in the interstitial region. These functions complement each other due to the projection technique.[42] The exchange and correlation potential is within the generalized gradient approximation as parametrized by Perdew, Burke, and Ernzerhof.[44] The weak van der Waals (vdW) interactions were computed within the so-called GGA-D3 approach developed by Grimme[45, 46] and later implemented in the VASP package.[47] To include the correlation effects of transition metal 3d electrons of Mn, we adopted the Dudarev *et al.* GGA+U method,[48] using a Hubbard U of 4 eV and an exchange parameter *J* of 1 eV. Note that in this GGA+U approach the exchange parameter is included in the Coulomb parameter and only the effective value $U_{eff} = U - J = 3$ eV is directly relevant. This value of $U_{eff}$ was found by Brumboiu *et al.*[49] to describe the HOMO-LUMO gap of MnPc and higher values of $U_{eff}$ did not change it (see Figure 2b of Ref. [49]). Other ab initio calculations also adopted the same effective Hubbard interaction for MnPc.[50-52]

The substrates are modeled by a slab consisting of 3 layers of 16 x 16 atoms per layer. We used 996 atoms in the supercell, including the 228 atoms of the four MnPc molecules, and more than 2 nm vacuum to the next periodic repeated slab. We should emphasis that 3 monolayers are enough for the convergence of the total energy and density of states. We showed that 3 and 5 monolayer supercells produced adsorption energies and DOS that are not significantly different from each other. We have checked that a kinetic-energy cutoff of 500 eV for the plane wave basis set and a charge-density cutoff of 8 times larger are sufficient for the convergence of the total energy. Because of the large number of atoms per supercell we used only a single Γ-point in the first Brillouin zone. For the calculation of the Fermi level we used Methfessel-Paxton method[53] with a smearing of 0.1 eV. The Kohn-Sham self-consistent total energy is converged to within $10^{-6}$ eV. For the atomic relaxation, the lowest layers of silver is fixed whereas both the MnPc molecules and the topmost two metallic layers are fully relaxed. The force on the atoms are converged within $10^{-3}$ eV $Å^{-1}$. To make sure our parameters and settings are reasonable, we checked that the density of states of the free MnPc molecule is in good agreement with existing results.[54] To compute the dissociation energy of the $(MnPc)_2$ dimer, we have computed the total energy of the dimer in the same supercell used for the calculation of the electronic structure. The dissociation energy is defined as the difference between the total energy of the dimer in its ground state and that of the two MnPc molecules taken separately.

## 4. CONCLUSION

In summary, we have shown that a magnetic $(MnPc)_2$ dimer-switch is a viable entity on a metal substrate. The approach goes beyond the simple 2D organization and explores the possibility of building molecular functional objects in 3D. Both, switching and recognition between a ferromagnetic and an antiferromagnetic state are achieved by means of the same STM device, upon a small and reversible conformational change of the dimer. The influence of the substrate and the origin of the intermolecular coupling have been carefully evaluated by means of GGA+U calculations. The two states of the switch also have distinct topographic and spectroscopic fingerprints which makes them easily identifiable. Therefore, our findings by means of STM and *ab-initio* calculation provide strong arguments for the use of magnetic dimers as bi-stable molecular switches and open potential research avenues towards applications in quantum materials and spintronics. No similar reversible switching has been observed yet on minimal units of only a few molecules, typically a dimer.

**Supporting Information**

Supporting Information is available from Wiley Online Library or from the authors.

**Acknowledgements**

This work of the Interdisciplinary Thematic Institute QMat, as part of the ITI 2021-2028 program of the University of Strasbourg, was supported by IdEx Unistra (ANR-10-IDEX-0002), by SFRI STRATUS project (ANR-20-SFRI-0012) and by ANR-11-LABX-0058_NIE under the framework of the French Investments for the Future Program. The computational work was performed using Strasbourg HPC resources and GENCI-CINES Grant 2021-gem1100.

**Conflict of Interest**

The authors declare no are no conflict of interest.

**Corresponding Author**

* E-mail: jean-pierre.bucher@ipcms.unistra.fr

**Present Addresses**

† Present Address: Peter Grünberg Institut, Forschungszentrum Jülich, Nordrhein-Westfalen, Germany.

**Table of contents entry**:

On-surface assembled dimers of Mn-Phthalocyanine $(MnPc)_2$ can be reversibly switched between an antiferromagnetic and a ferromagnetic state upon a small conformational change controlled by STM manipulation. The two states of the molecular switch are easily identified spectroscopically by their Kondo resonance, opening interesting routes for applications in molecular spintronics, logics and computing.

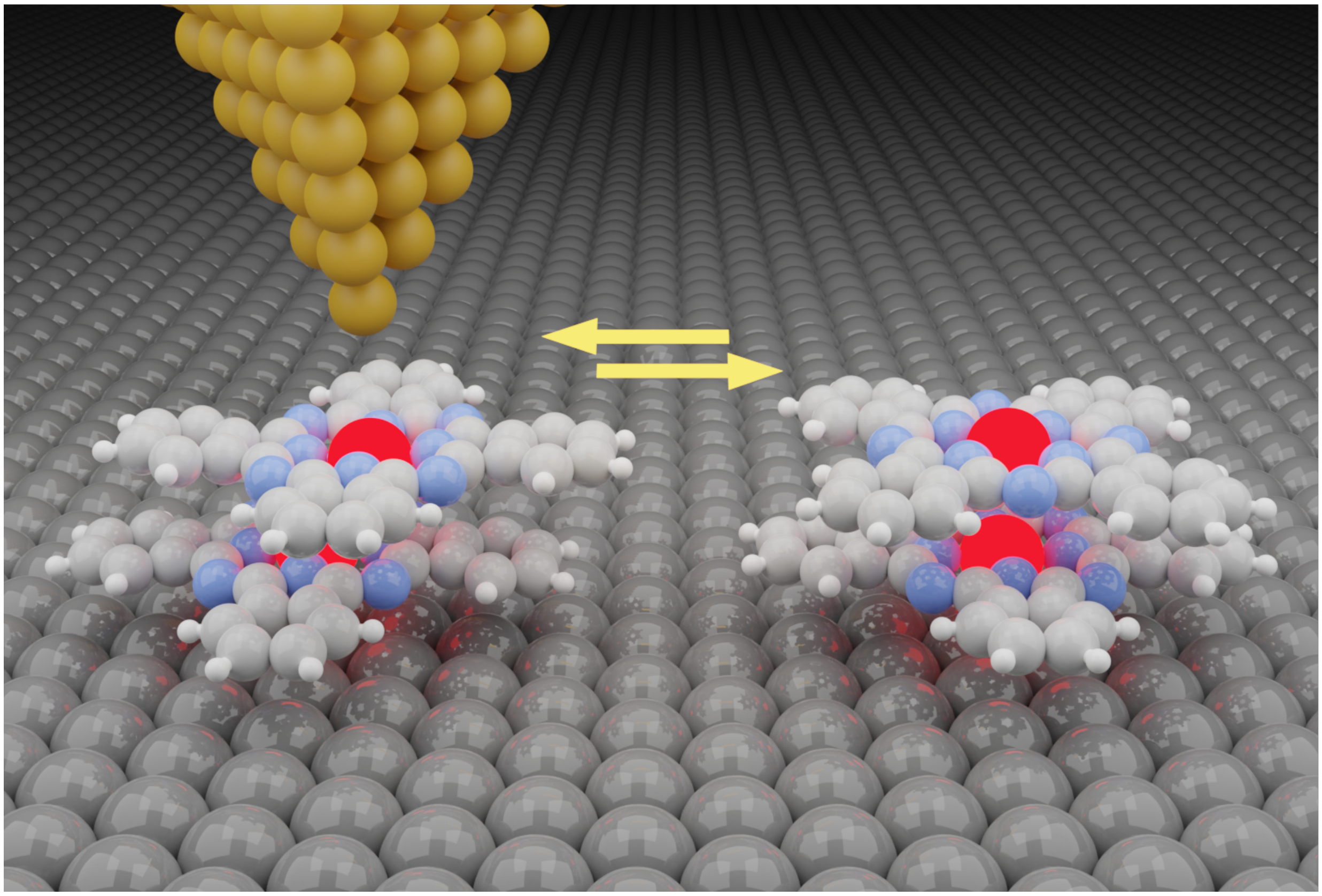

Supporting Information

# When Molecular Dimerization Induces Magnetic Bi-Stability at the Metal-Organic Interface

*Rouzhaji Tuerhong, Franck Ngassam, Mebarek Alouani, Jean-Pierre Bucher**

Université de Strasbourg, IPCMS UMR 7504, 67034 Strasbourg, France

*E-mail: jean-pierre.bucher@ipcms.unistra.fr

**Table of content**

## S1. Stacking geometries in MPc bulk materials

It is of interest to compare our results to the situation encountered in thick layers or bulk metal-phthalocyanine (MPc) materials. In large systems, the MPc molecules generally form columnar structures where columns are formed by stacks of planar molecules parallel to each other with the central metal ions forming the axis of the column. Bulk MPc crystals may exhibit two monoclinic columnar structures with stack angles between molecular planes and column axis of 65° and 45° for α- and β-phases, respectively. The β-phase of MnPc molecules is mostly found in bulk crystals and is the most stable crystal structure, whereas the meta-stable α-phase is often found in thin films. For MnPc molecules in both α- and β-phases, the distance between parallel molecular planes is $d_z = 3.4 \pm 0.1$ Å [1-3]. In the α-phase, adjacent layer MnPc molecules are shifted with respect to each other by $|d_{xy}| = 1.54$ Å along the in-plane (110) axis, so that the Mn center of a MnPc is positioned at the center of the pyrrolic rings of the two adjacent layer MnPc molecules [1,2]. In the β-phase, the molecules shift along the in-plane (110) direction by $|d_{xy}| = 3.4$ Å, so that the Mn ion of one molecule is located directly above or below the N atom bridging two isoindole units of the adjacent MnPc molecule [3].

Here, the STM observation shows that the structures adopted by second layer MnPc molecules are different from both, α- and β-phase. The two dimer structures are shown in Figure 1 (main text). In the *shifted* configuration of MnPc-dimer, the inter-layer separation $d_z = 3.0$ Å is significantly shorter than that of α- and β-structures. Additionally, the directional displacement and stack angle of the *shifted* MnPc-dimer are different from those of α- and β-phase. In the *shifted* MnPc dimer, the Mn atom is located above the first shell N atom along (100) axis of the adjacent MnPc molecule. Finally, in our *on-top* configuration, the two molecules in the $(MnPc)_2$ dimer are rotated by 45˚ with respect to each other around the z-axis that runs through the two vertically aligned the Mn atoms that have an inter-layer separation of $d_z = 3.45$ Å.

In addition to the bulk phase of MnPc mentioned above, similar columnar stacks of metallo-Pc molecules has been observed in the multilayer films of FePc grown on Au(111) [4] and CoPc on Pb(111) [5], however no emphasis was placed on dimers. In the case of CoPc/Pb(111) [5] the Kondo resonance was reported to be absent for the first CoPc layer due to the quenching of the spin moment upon adsorption on the metallic Pb(111) substrate. A modest Kondo screening appeared however $T_K < 10\ K$ for a second layer "on-top" CoPc molecule, despite the decoupling between the second layer CoPc molecule and the Pb(111) substrate. Additional layers were found to stack in the shifted fashion. This situation is very different from the case MnPc molecule on Ag(111) where the delocalized $\pi$-electrons of the ligand produce an unusual extension of the Kondo resonance over first layer MnPc molecules leading to large Kondo temperatures [6].

### S2. Manipulation and switching by means of the STM tip

**Lateral manipulation and molecular dimer formation.** To investigate possible stacking configuration of a MnPc dimer, the STM tip is used to manipulate the isolated single molecules in the second layer and positioning it at a preferential adsorption site above a first layer MnPc molecule (see Figure S1). Single MnPc molecules have been extracted from the edge of a second layer domain by the lateral manipulation with STM tip. The lateral manipulation of a single MnPc molecule is performed as follows:

(i) A topography image is aquired above an area comprising a first monolayer of MnPc molecules toped by a second layer molecular island at 10 pA and -0.65 V. As seen from Figure S1a, the upper molecular domain and the lower monolayer are clearly resolved down to the sub-molecular scale.

(ii) The STM tip is then positioned above a MnPc molecule (see the circle in Figure S1a) at the edge of second layer domain and the tip height is decreased until the tunneling current reaches a set point of 1 nA and +0.5 V. These conditions constitute the proper tunneling parameters for the lateral manipulation of a MnPc molecule.

(iii) The feedback loop is disabled and the STM tip is moved laterally along a predefined path (see the arrow in Figure S1a) until it reaches the desired position on the first MnPc monolayer. Simultaneously, the induced variation of the tunneling current is recorded as a function of lateral displacement of the STM tip (see Figure S1c).

(iv) The feedback loop is turned on again and the STM tip is retracted to the imaging height where the molecule-tip interaction is negligible as in (i).

(v) Finally, the topography image of the same surface area is recorded with the same parameters as in (i) to confirm whether the molecule has moved to the desired new position (see Figure S1b).

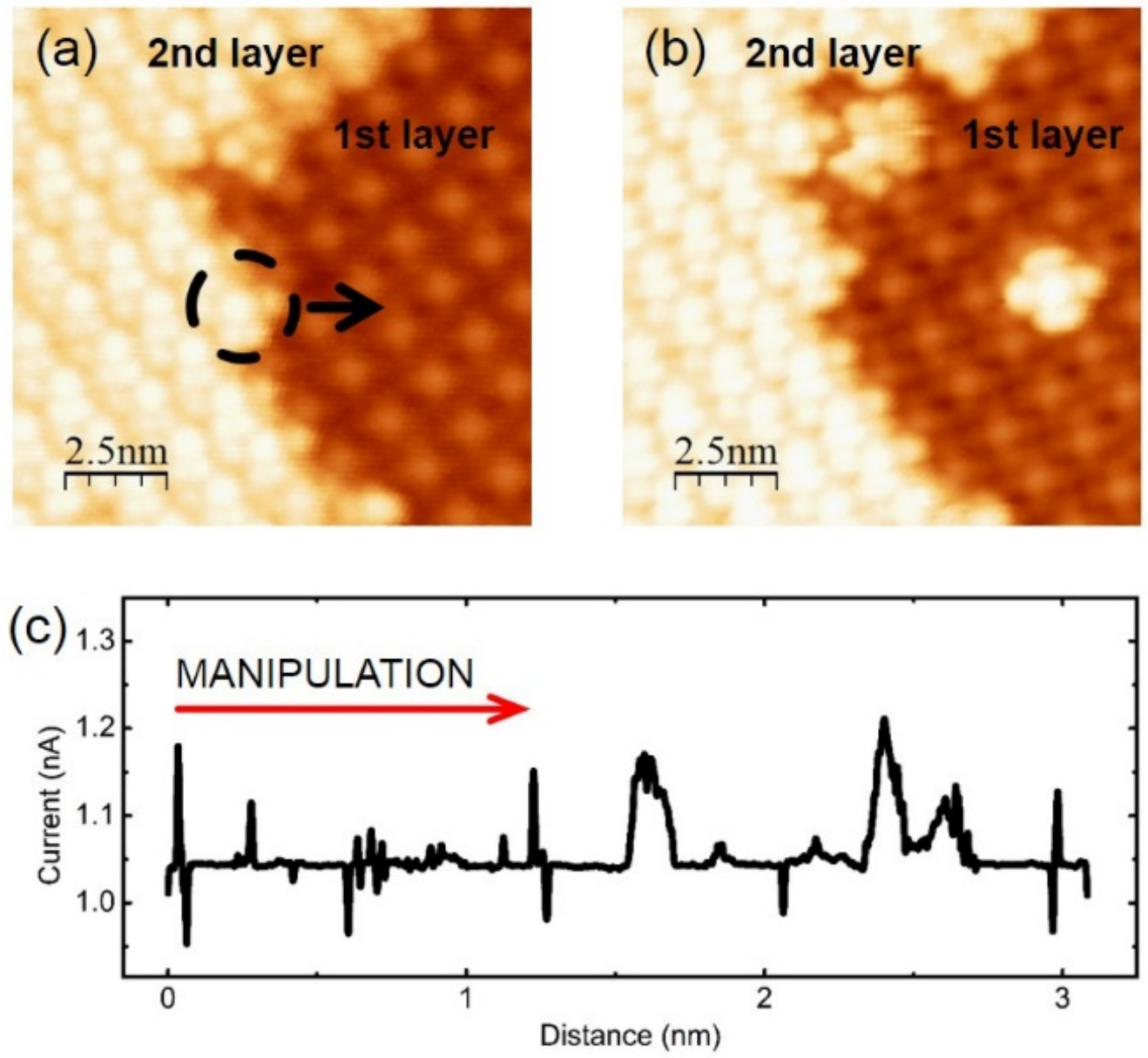

**Figure S1**. STM images of (a) before and (b) after the extraction of a single MnPc molecule from the edge of second layer domain and positioning at a preferred adsorption site on the monolayer structure using lateral manipulation with the STM tip (Imaging parameter for (a) and (b): -0.65 V, 10 pA) T = 4.5 K. (c) Variation of the tunneling current during the lateral manipulation of a MnPc molecule (set point: 0.5 V, 1 nA).

The outcome of lateral manipulation following steps (i)-(v) is summarized in Figure S1 and confirms that a single MnPc molecule is successfully extracted (Figure S1a) and moved to a "shifted" adsorption site above a molecule in the first monolayer (Figure S1b).

**Conformational switching of molecular dimers.** Here we further demonstrate the conformational switching of the MnPc dimers induced by the controlled manipulation of a second layer molecule by means of the STM tip. The *on-top* and *shifted* configurations of MnPc dimers are stable under the STM measurements providing non-invasive tunneling parameters are used for the imaging. We observed however that it is possible to manipulate the dimer conformation at some critical tunneling parameters, thus controlling the adsorption site of a single MnPc molecule on top of a first molecular layer. Such critical conditions induce an interaction between the STM tip and the molecule that helps manipulate the second layer MnPc molecule. Here, we first demonstrate the conformational switching of MnPc dimer during controlled incrementation of the tunneling current at low negative sample bias.

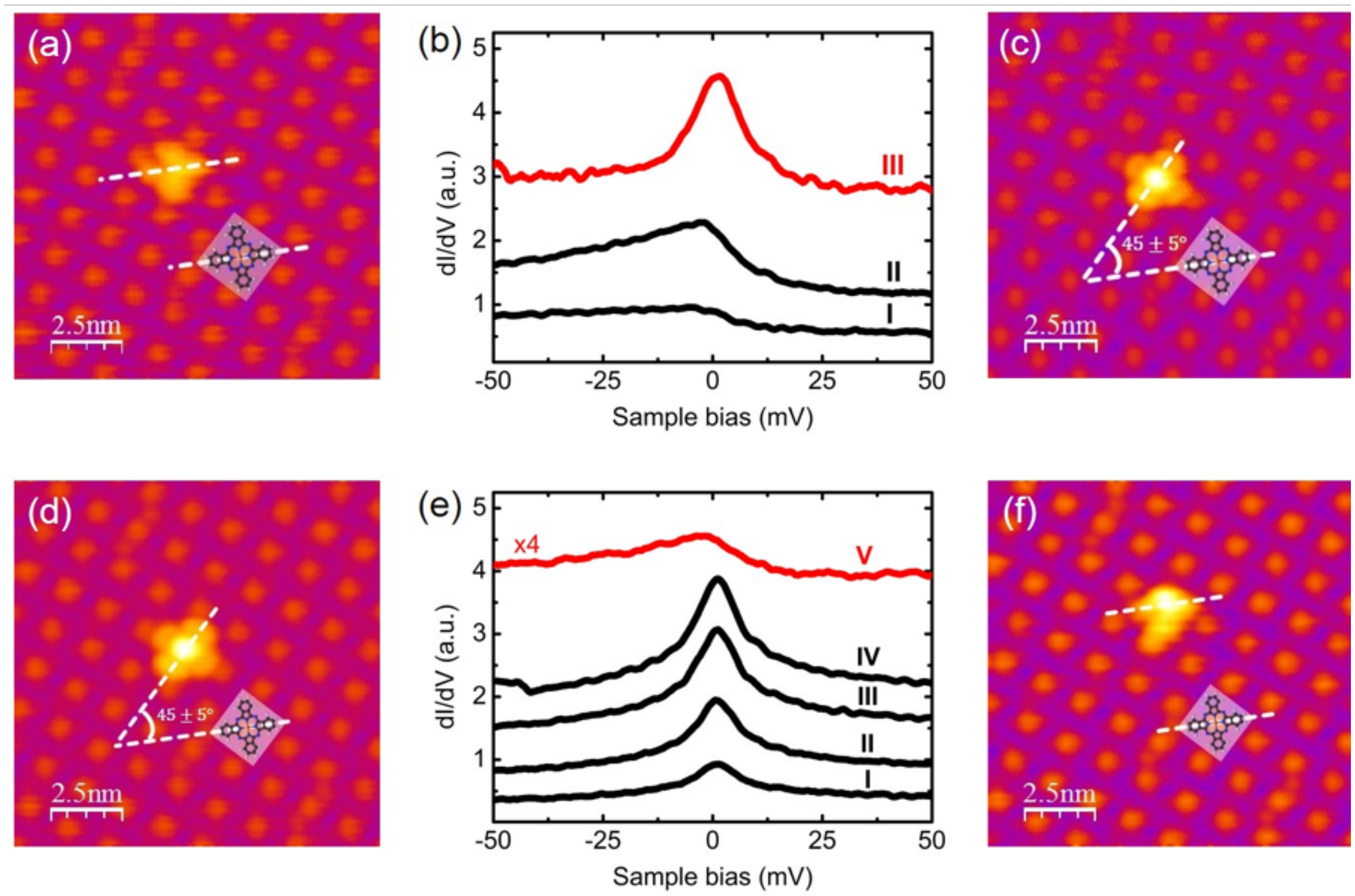

**Figure S2.** The conformational switching of single MnPc molecule on top of the first MnPc monolayer on Ag(111) surface. (a) STM image of a single MnPc molecule in the second layer with *shifted* configuration (-0.5 V, 10 pA). (b) Tunneling spectra measured above the Mn center of the upper MnPc molecule in (a), by increasing the tunneling current set point from (I) 100 pA to (II) 200 pA and up to (III) 300 pA while keeping the sample bias voltage constant at -0.1 V. A dramatic evolution is observed in spectrum (III) as signified by the narrowing of the zero-bias peak following a step-like increase of the conductance change at negative bias. (c) STM image of the same molecule immediately after the measurement of spectrum (III), acquired with the same parameters as in (a). Now the MnPc molecule has switched to the *on-top* configuration. The lower part is imaged with -0.5 V, 50 pA to resolve the orientation of MnPc molecules in the first monolayer. (d) STM image of the same molecule as in (c) with *on-top* configuration (-0.5 V, 10 pA). (e) Tunneling spectra measured above the Mn of the *on-top* molecule in (d), by increasing the tunneling current set point from (I) 100 pA, (II) 200 pA, (III) 300 pA, (IV) 400 pA and up to (V) 500 pA at constant the sample bias of -0.1 V. When the tunneling current set point is increased, a spectacular evolution is observed for spectrum (V) which bears the imprint of the *shifted* configuration as revealed by the drop and widening of the ZBP. (f) STM image of the molecule immediately after the measurement of spectrum (V), and acquired with the same parameters as in (d), confirming the switching to the *shifted* configuration. It is noticeable though that the molecule has simultaneously jumped to the next neighboring adsorption site (compare with (a)).

It proceeds in three steps as summarized in Figure S2. First, an STM image is acquired at 10 pA and -0.5 V (Figure S2a). The parallel molecular orientation of the upper MnPc molecule with respect to the the first molecular layer reveals a *shifted* configuration. Next step (Figure S2b) consists in keeping the STM tip above the central Mn ion of the upper MnPc molecule and increasing the tunneling current set point from 100 pA, to 200 pA and 300 pA. At each set point, the feedback loop is disabled and a conductance spectrum is recorded. The significant increase of the tip-molecule interaction from I to III leads to a dramatic change in spectrum III that shows a step in the conductance at negative bias accompanied by a significant narrowing of the Kondo resonance that is the signature of the *on-top* dimer. Finally, the topographic confirmation that the dimer has switched from the *shifted* to the *on-top* configuration is obtained in Figure S2c by imaging again the surface with the same scan parameters as in Figure S2a. The reversible switching is also possible as described in Figures S2d to S2f. Our spectroscopic approach thus constitutes a real time observation of the molecular switching. The success rate of reversible switching is about 60%.

Occasionally we could also observe a configuration switching during STM imaging. We found that the stability of an adsorption site of the upper molecule during the scanning strongly depends on the polarity and magnitude of the applied sample bias voltage. A small and negative sample bias stabilizes the MnPc dimer configurations, whereas a large negative or positive bias may induce a motion of the upper molecule along with the STM tip during the imaging (see lateral manipulation above). In some cases, the molecule is picked-up by the STM tip.

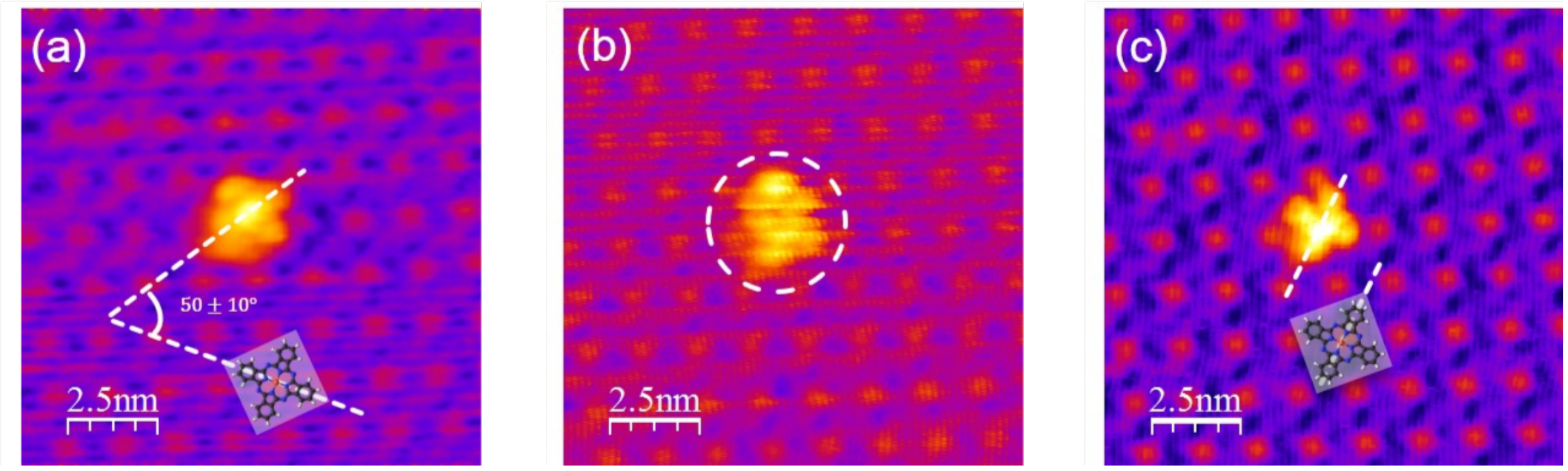

**Figure S3.** Conformational switching of single MnPc molecule. (a) A single MnPc molecule is adsorbed in the *on-top* configuration of a first layer of MnPc molecule as highlighted by the azimuthal rotation. This *on-top* MnPc configuration is stabilized at scan parameters: -0.5 V, 10 pA. (b) STM image of the same molecule than (a) imaged with different scan parameters: -0.8 V, 10 pA. A conformational switching of the molecule is induced between two states during the scan as evidenced by the persistent image inside the circle. (c) Same molecule as before, imaged with the non-invasive parameters used in (a). It is found that the molecule has switched into the *shifted* configuration during the scan in (b).

As an example, Figure S3 displays a conformational switching of a MnPc molecule on-top of the MnPc monolayer on Ag(111) surface. Figure S3a, shows an STM image acquired at $-0.5\,V$ and $10\,pA$, which is below the switching threshold, i.e. the bias voltage is sufficiently low to stabilize the single MnPc molecule in the second layer. The molecular orientation highlighted by the superposed structure on the first molecular layer and the measured azimuthal angle of $\theta = 50 \pm 10°$ reflects the fact that the upper MnPc molecule is stabilized in the *on-top* configuration above the MnPc monolayer (see Figure S3a). Then, by scanning this surface area at a bias voltages of $-0.8\,V$ and $10\,pA$, we observed a conformational switching of the same MnPc molecule between two states (persistent image inside the circle) in Figure S3b. Finaly, Figure S3c shows the same surface area scanned under the same non-invasive conditions as before. It is observed that the MnPc molecule in the second layer is stable again but has switched to the *shifted* configuration.

**S3. Calculated values from GGA+U**

The results of the magnetic properties of the two MnPc molecules in vacuum and adsorbed on the Ag(111) surface are presented in Table S1. Here we present both the ferromagnetic (FM) and antiferromagnetic (AFM) configurations and when the two molecules are on-top of each other or when they are shifted with respect of each other. In the case of on-top configuration, the Mn are on-top of each other but the Pc of the upper molecule is rotated by 45° with respect to that of the lower molecule. In the case of the shifted configuration, the Mn atom of the upper molecule is on top of the one of the first neighbor N atom of the Mn atom of the lower molecule.

As it is shown in Table S1, when the molecules are on-top of each other, the most stable magnetic configuration is antiferromagnetic (AFM), mainly due to a direct hybridization of the $d_{z^2}$ Mn orbitals, whereas in the shifted geometry the coupling is ferromagnetic (FM), due to the mediation by the nitrogen atom of the first molecule, thus to an indirect superexchange. When the molecules are adsorbed on Ag(111), see Table S2, the magnetic state remain unchanged albeit a reduced strength of the magnetic coupling appears. It is also interesting to notice that the distance between the two molecules is reduced for the shifted configuration due to the reduced electrostatic repulsion between the Mn atoms of the two molecules. Notice also that the magnetic moment on the Pc part of the molecule is opposed to that of Mn.

**Table S1** Ab-initio calculations for the *on-top* and *shifted* geometries in vacuum. Subscript 1 and 2 refer to the first and second molecule of the dimer.

| in Vacuum | **On-top** | **Shifted** |
|---|---|---|
| Magnetic state | *AFM* | *FM* |
| *Mn moment* ($\mu_B$) | $m_1 = -m_2 = 3.40$ | $m_1 = m_2 = 3.52$ |
| *Pc moment* ($\mu_B$) | $m_1 = -m_2 = -0.37$ | $m_1 = m_2 = -0.42$ |
| *coupling: J (meV)* | *-38* | *+122* |
| *distance:* $d_{MnPc1\text{-}MnPc2}$ *(pm)* | *335* | *278* |

**Table S2** Ab-initio calculations for the *on-top* and *shifted* geometries adsorbed on Ag(111). Subscript 1 and 2 refer to the lower and upper molecule with respect to the substrate.

| on Ag(111) | **On-top** | **Shifted** |
|---|---|---|
| Magnetic state | *AFM* | *FM* |
| *Mn moment* ($\mu_B$) | $m_1 = 3.42$<br>$m_2 = -3.47$ | $m_1 = 3.56$<br>$m_2 = 3.56$ |
| *Pc moment* ($\mu_B$) | $m_1 = -0.61$<br>$m_2 = +0.36$ | $m_1 = -0.51$<br>$m_2 = -0.42$ |
| *coupling: J (meV)* | *-11* | *+56* |
| *distance:* $d_{MnPc1\text{-}MnPc2}$ *(pm)* | *341* | *290* |

## S4. Spatial distribution of the spin density in vacuum

Figure S1 shows top and side views of the spatial distribution of the magnetization of the two (on-top and shifted) configurations when the dimers are in vacuum. It shows no significant difference compared to the dimer adsorbed on Ag(111) of Figure 2 although, as shown in Table S1, numerical values of the coupling can be significantly different. Figure S4a clearly shows an AFM coupling whereas Figure S4b shows the FM coupling of the two molecules of the dimer. It is also clear that the Pc parts of both molecules are coupled antiferromagnetically to the Mn atoms as is shown by the red color (spin up) and the blue color (spin down).

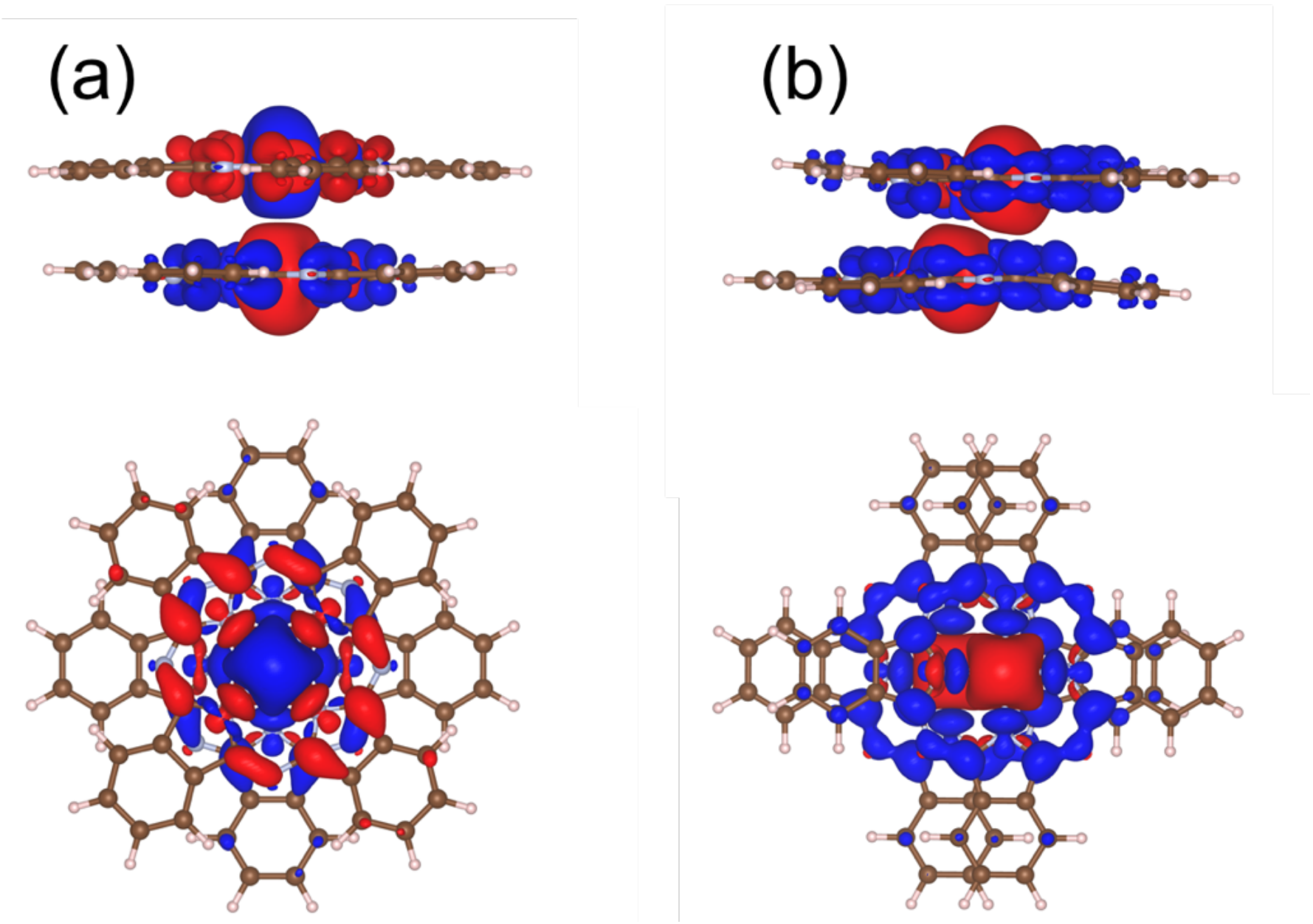

**Figure S4**. The spatial distribution of spin density in (a) On-top and (b) Shifted configuration of the MnPc dimer in the vacuum. The on-top configuration of the MnPc dimer couples antiferromagnetically (AFM) through direct exchange, and switches to the ferromagnetic (FM) coupling through superexchange in the shifted configuration. The red and blue colors are representing the spin-up and spin-down density, respectively. The isosurface value is 0.02 e/Å$^3$.

## S5. Site selective PDOS and occupation of Mn 3d orbitals

In the following, density of states projected on Mn $d_{xy}$, $d_{x^2-y^2}$, $d_{xz/yz}$, and $d_{z^2}$ orbitals (PDOS) are analyzed for the Mn1 and Mn2 atoms of both, the on-top and the shifted dimer in vacuum and on Ag(111). In order to better understand the magnetic coupling mediated by the nitrogen, the projection onto the p-orbitals of specific N-atoms has been computed as well.

**On-top dimer.** The Mn LDOS of the vacuum dimer on Mn1 and Mn2 are identical and opposed in spin direction (Figure S5a). Individually, they are not significantly different from the monomer LDOS. Although significant changes appear when the dimer is contacted with the surface, see Figure S5b, the Mn1 LDOS comes close to the one of the contacted monomer whereas the Mn2 LDOS is almost identical to the vacuum dimer. Overall the AFM character is retained but whereas Mn2 shows an almost undisturbed $d_{z^2}$ state, Mn1, in contact, shows a strong reduction of the $d_{z^2}$ state at -4 eV that splits into multiple peaks, similar to the monomer in contact with the substrate.

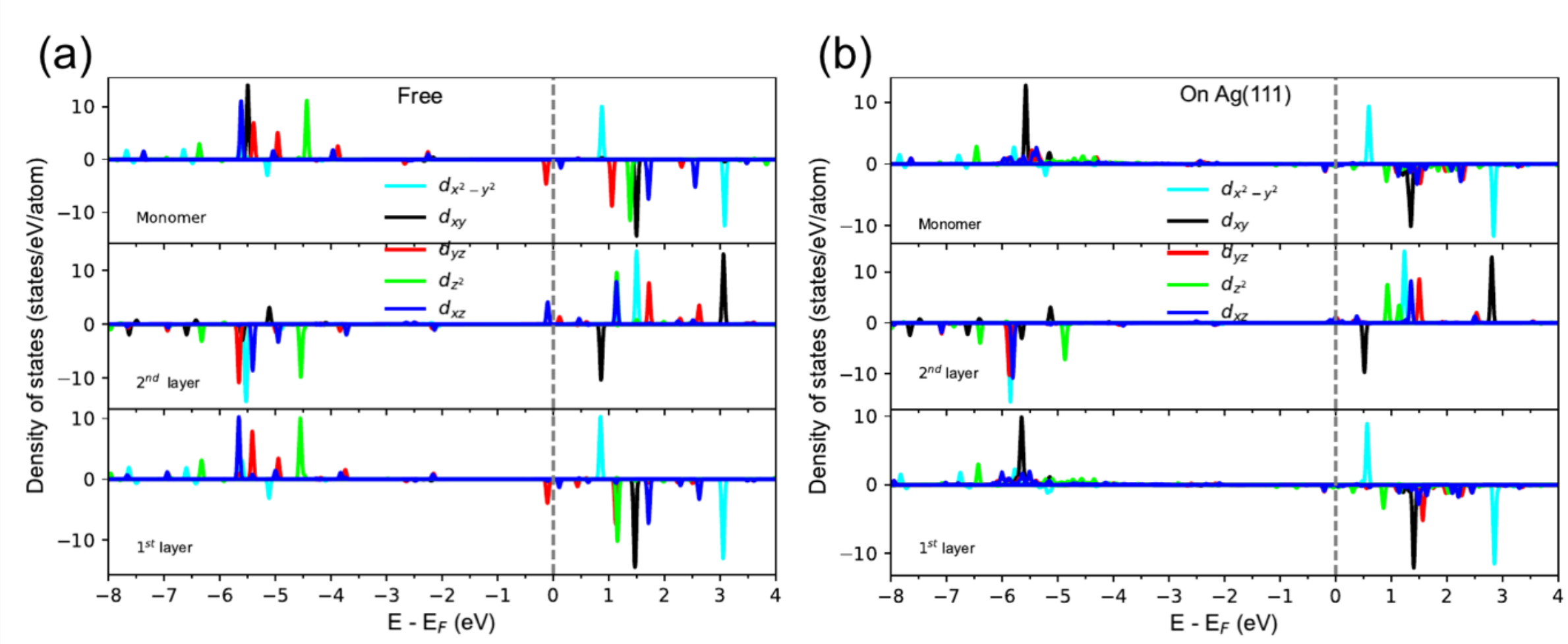

**Figure S5**. PDOS for the "on-top" dimer projected on the d-states of either Mn1 or Mn2. (a) Vacuum (b) on Ag(111). The monomer in the same situation (top of the panel) is shown as well for comparison.

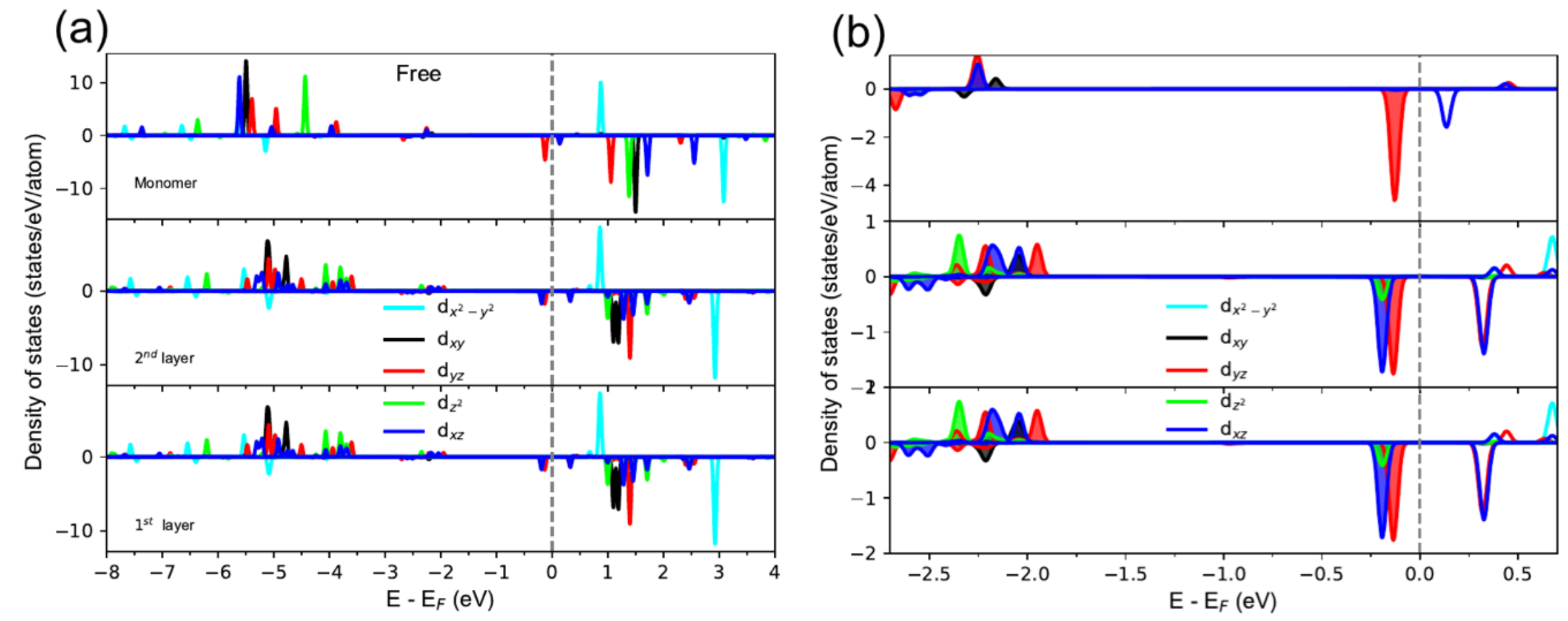

**Figure S6**. PDOS for the "shifted" dimer projected on the d-states of either Mn1 or Mn2 in vacuum. (a) Wide range (b) Zoom. The monomer in the same situation is shown as well for comparison.

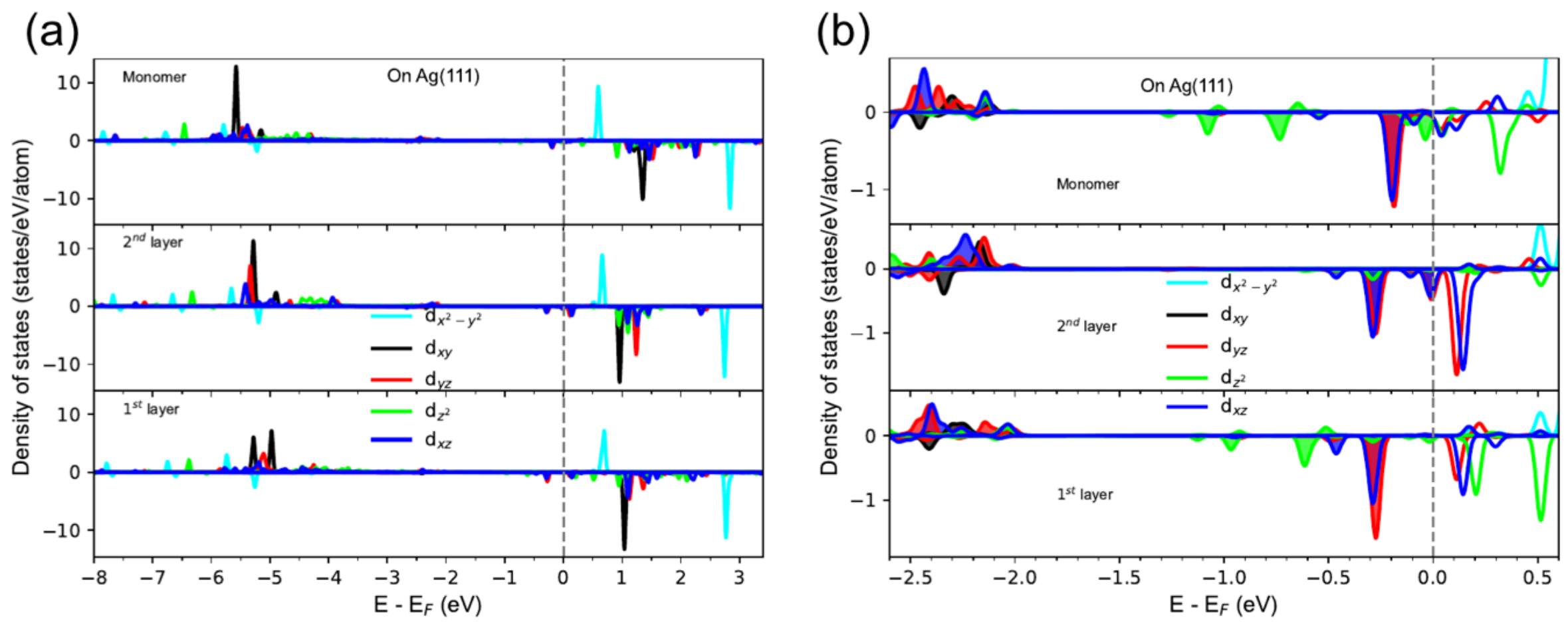

**Figure S7**. PDOS for the "shifted" dimer projected on the d-states of either Mn1 or Mn2 adsorbed on Ag(111). (a) Wide range (b) Zoom. The monomer in the same situation is shown as well for comparison.

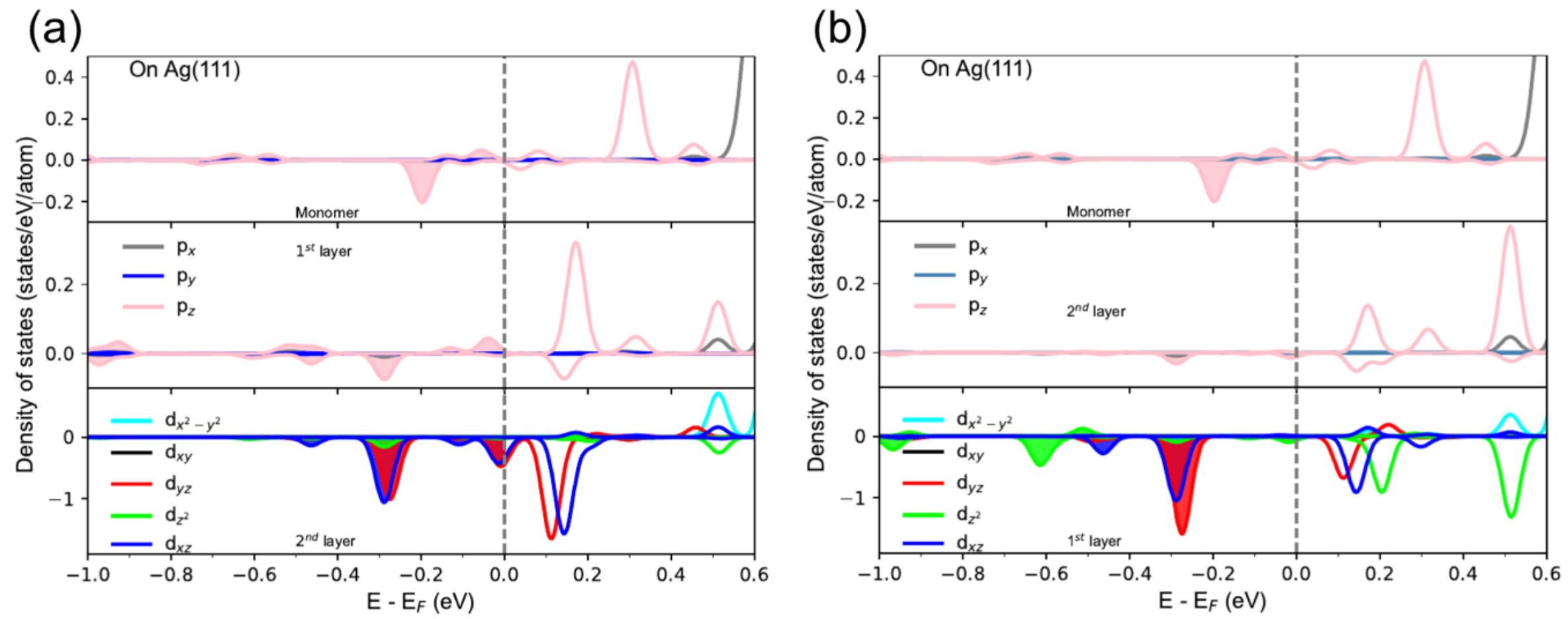

**Figure S8**. PDOS for the "shifted" dimer. Hybridization between p and d (near Fermi). (a) p-states of N1 and d-states of Mn2 (b) p-states of N2 and d-states of Mn1. The p-states of the monomer on Ag(111) are shown as well for comparison.

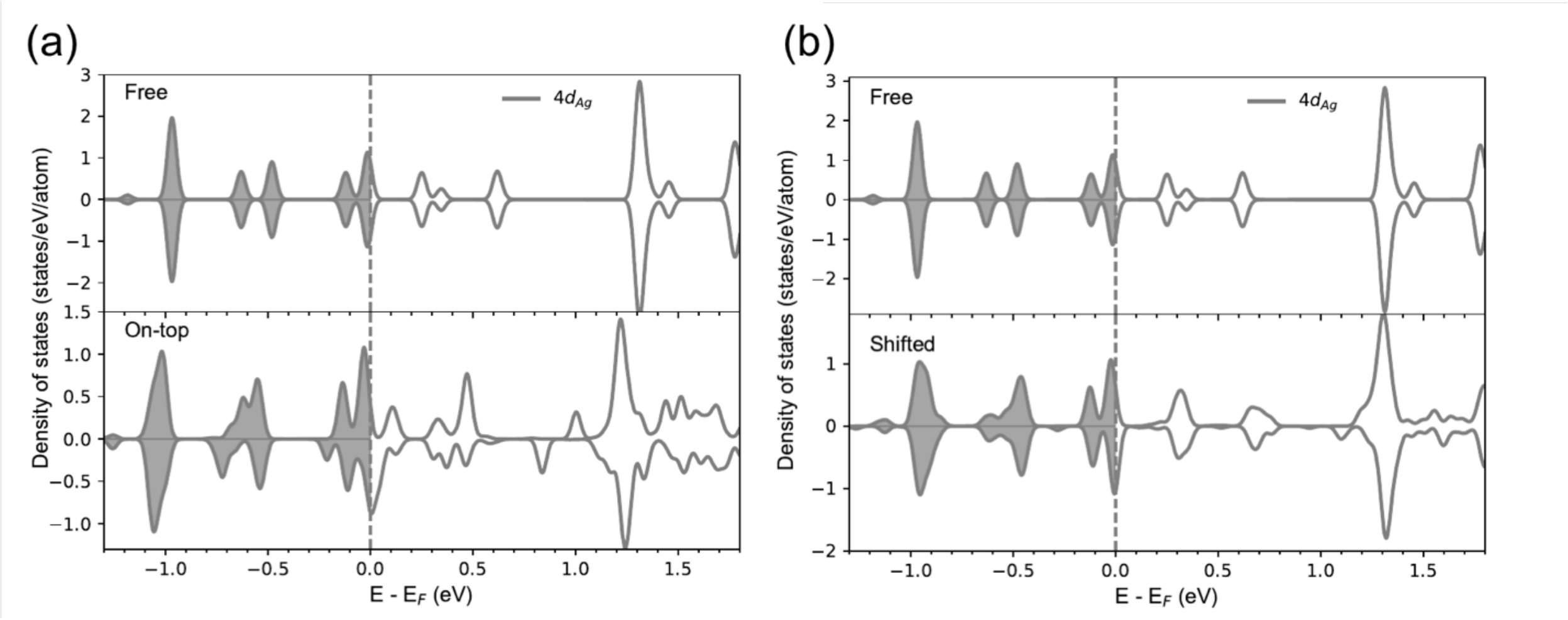

**Figure S9**. Polarisation of the substrate atoms by (a) on-top and (b) shifted dimer. Upper panels, prestine Ag top layer PDOS. Lower panel, PDOS on the silver atom just underneath the Mn of the $(MnPc)_2$ complex.

**Shifted dimer.** Contrary to the on-top case, the LDOS on Mn1 and Mn2 of the shifted dimer in the vacuum already show important changes compared to the monomer (Figure S6). Significant rearrangements of all orbitals are observed, out of plane as well as in plane for the reasons discussed before (Wannier analysis), in particular the $d_{z^2}$ peak close to -4 eV is now split in two but also $d_{xz}$ and $d_{yz}$ that were degenerate in the free monomer but are split in the dimer (see Figure S6a and b). When the shifted dimer is put in contact with the surface a new LDOS appears that is comparable on Mn1 and Mn2 but that does not compare with the monomer in contact nor with the vacuum situation (see Figure S7a and b). This reveals a higher symmetry between Mn1 and Mn2 and shows that the shifted dimer is less affected by the substrate than the on-top dimer.

Due to important role played by the N atoms in the shifted case, the hybridization between N-*p*-orbital and Mn-*d*-orbitals is now analyzed in more details for the shifted dimer adsorbed on Ag(111). Figure S8a and b show the near-Fermi LDOS on the N atom of one molecule and on the Mn atom of the other molecule. They show the important hybridization between N-$p_z$ orbitals and all the Mn-*d*-orbital with a z-contribution. In addition, it is clear that an in-plane interaction takes place along the N-Mn bond pertaining to the same MnPc molecule of the dimer.

Finally, the weaker interaction of the *shifted* compared to the *on-top* dimer with the Ag(111) substrate is confirmed by the LDOS of Figure S9 that shows a reduced polarization of the silver atoms topped by the shifted-dimer.

**Occupation of the Mn 3d orbitals.** The GGA+U calculation for the two configurations, provides information on d-orbital occupation for both spin states (Table S3). As explained in the article, the screening by the itinerant electrons of the substrate mainly concerns d-orbitals with a z-component among which the singly occupied $d_{z^2}$ is the prominent feature. In particular it is found that the $d_{x^2-y^2}$ and $d_{xy}$ orbitals have negligible Kondo sreening due to either a negligible polarization or an unfavorable symmetry.

**Table S3** Occupation of Mn 3d orbitals extracted from DFT calculations for the *on-top* and *shifted* configuration. Mn1 and Mn2 refer to the lower and upper molecule respectively.

| On-top \ Orbitals | | | $d_{xy}$ | $d_{yz}$ | $d_{z^2}$ | $d_{xz}$ | $d_{x^2-y^2}$ |
|---|---|---|---|---|---|---|---|
| In Vacuum | Mn1 | $n_\uparrow$ | 0.94 | 0.93 | 0.90 | 0.92 | 0.46 |
| | | $n_\downarrow$ | 0.03 | 0.05 | 0.05 | 0.37 | 0.31 |
| | Mn2 | $n_\uparrow$ | 0.31 | 0.21 | 0.05 | 0.20 | 0.03 |
| | | $n_\downarrow$ | 0.46 | 0.92 | 0.90 | 0.92 | 0.94 |
| On Ag(111) | Mn1 | $n_\uparrow$ | 0.94 | 0.93 | 0.89 | 0.92 | 0.46 |
| | | $n_\downarrow$ | 0.03 | 0.14 | 0.17 | 0.15 | 0.30 |
| | Mn2 | $n_\uparrow$ | 0.31 | 0.17 | 0.05 | 0.19 | 0.03 |
| | | $n_\downarrow$ | 0.47 | 0.93 | 0.90 | 0.93 | 0.94 |

| Shifted \ Orbitals | | | $d_{xy}$ | $d_{yz}$ | $d_{z^2}$ | $d_{xz}$ | $d_{x^2-y^2}$ |
|---|---|---|---|---|---|---|---|
| In Vacuum | Mn1 | $n_\uparrow$ | 0.94 | 0.93 | 0.92 | 0.93 | 0.47 |
| | | $n_\downarrow$ | 0.03 | 0.16 | 0.09 | 0.15 | 0.31 |
| | Mn2 | $n_\uparrow$ | 0.94 | 0.93 | 0.92 | 0.93 | 0.47 |
| | | $n_\downarrow$ | 0.03 | 0.16 | 0.05 | 0.15 | 0.31 |
| On Ag(111) | Mn1 | $n_\uparrow$ | 0.94 | 0.93 | 0.91 | 0.93 | 0.47 |
| | | $n_\downarrow$ | 0.03 | 0.15 | 0.12 | 0.13 | 0.31 |
| | Mn2 | $n_\uparrow$ | 0.94 | 0.93 | 0.92 | 0.93 | 0.48 |
| | | $n_\downarrow$ | 0.04 | 0.14 | 0.07 | 0.15 | 0.31 |

## S6. Wannier function representation projected to atomic orbitals

The Wannier function representation was used in the main text (Figure 3) to give a detailed analysis of the magnetic coupling of the $(MnPc)_2$ dimers adsorbed on Ag(111). The corresponding vacuum situation is represented in Figure S10 for both, the on-top and the shifted dimers. The side and top views of the Wannier functions show explicitly their positive and negative parts, allowing a better understanding of the hybridization of the orbitals. Although in the vacuum the symmetry between the two molecules of the dimer is preserved, the discussion in terms of orbital couplings remains the same as in the adsorbed case. In the on-top configuration of Figure S10a the two $d_{z^2}$ orbitals of the Mn atoms are the only ones that strongly hybridize which explains the AFM coupling of the two molecules. The shifted configuration of Figure S10b, on the contrary, shows that the Mn atom of the one molecule and the N atom of the opposite molecule are likely to form an antibonding state which leads to a FM coupling of the two Mn by an indirect superexchange.

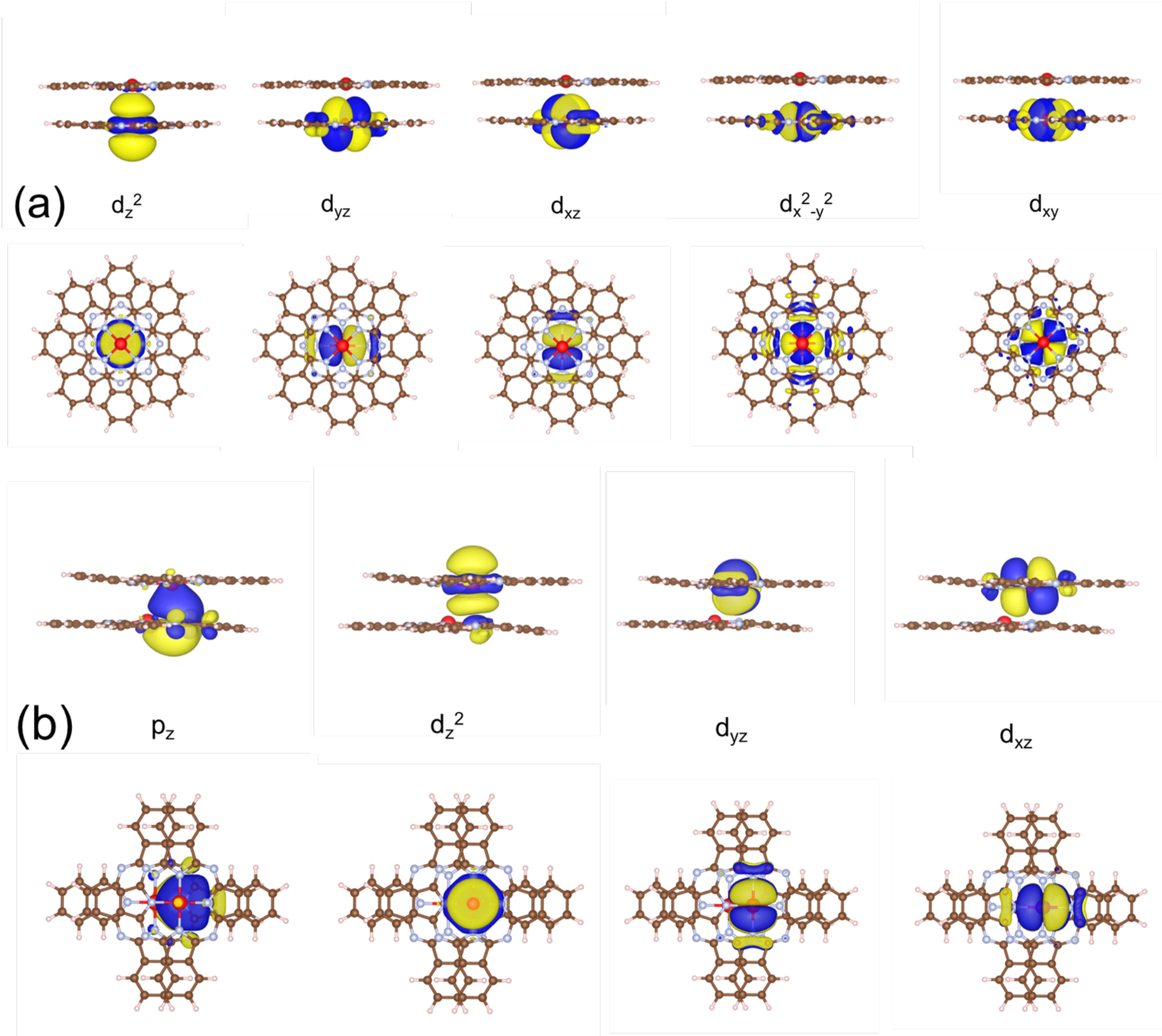

**Figure S10**. Wannier function analysis for the $(MnPc)_2$ dimer in the vacuum for comparison with $(MnPc)_2$/Ag(111) of Figure 3 of the main text. (a) On-top dimer with lower layer Wannier function projected onto the d-states. Significant overlap is only possible for $d_{z^2}$ orbitals since other orbitals are less extended towards the z-direction, but also for symmetry reasons in the case of $d_{xz}$ and $d_{yz}$ orbitals since the 2nd layer orbitals of this kind are is rotated by 45° with respect to the 1st layer thus strongly reducing the overlap. (b) Shifted dimer. The N atom of the lower molecule and the Mn atom of the upper molecule are forming an antibonding state which leads to a FM coupling of the two Mn via an indirect superexchange.

## S7. Temperature-dependence of dI/dV spectra and Kondo temperature

As discussed in the main text, the ZBP can be well fitted by a Fano lineshape, which constitutes the best match to our data [7]. To confirm that the observed ZBP is due to the Kondo resonance (Fermi liquid behavior), we measured a series of dI/dV spectra at temperatures ranging from 4.5 K up to 100 K. The characteristic temperature dependent behavior of the Kondo resonance can then be determined by carrying out a correction of the spectra at each temperature to remove the effect of the modulation voltage by means of the following relation:

$$\Gamma_F = \sqrt{(1.7eV_m)^2 + (2\Gamma_K)^2}/2 \quad , \tag{1}$$

where $\Gamma_F$ is the half-width at half maximum (HWHM) of the Fano lineshape, $V_m$ is the modulation voltage (3–5 mV peak-to-peak in this work), and $\Gamma_K$ is the intrinsic, corrected HWHM of the Kondo resonance. For a set of spectra measured at different experimental temperatures, the results can be summarized in a $\Gamma_K$ versus $T$ diagram as shown in Figure 4f that fits the Fermi-liquid variation [8,9] :

$$\Gamma_K = \sqrt{(\alpha k_B T)^2 + (2k_B T_K)^2}/2 \quad , \tag{2}$$

thus confirming the Kondo nature of the observed ZBP. As can be seen, $\Gamma_K$ grows linearly at high temperatures where the slope $\alpha$ can be determined, but saturates at a finite value at low temperature (solid red line in Figure 4f). The $\Gamma_K$ value at $T = 0K$ defines the Kondo temperature $T_K = \Gamma_K/k_B$, where $k_B$ is the Boltzmann constant. As an example, for a single MnPc molecule on Ag(111) we find $T_K = 130 \pm 10\ K$ and a parameter $\alpha = 5.85 \pm 0.10$.

In order to extract the Kondo temperature $T_K$ for the MnPc dimers, the ZBP are first fitted by a Fano function (Figure 4c and d) to extract $\Gamma_F$. By eliminating the thermal broadening and modulation voltage broadening, using equation (1) and (2), and assuming $\alpha = 5.85 \pm 0.10$ for the MnPc dimers as well, we get $\Gamma_K = 5.2 \pm 0.5\ meV$ and $10.8 \pm 1\ meV$ for the *on-top* and *shifted* dimers respectively and consequently $T_K = 60 \pm 4\ K$ and $125 \pm 7\ K$ for the Kondo temperatures.

## S8. Wide range dI/dV spectra

The spectrum above the second layer MnPc molecule in Figure S11b shows a large and broad feature at +1.4 eV as well as much weaker features below $E_F$, in the range under consideration, that compares well with the unoccupied and occupied states of the Mn 3d orbital in the first- and second-layer LDOS. As discussed in the main paper, the zero-bias peaks in (c) are signatures of the Kondo screening of Mn 3d spin.

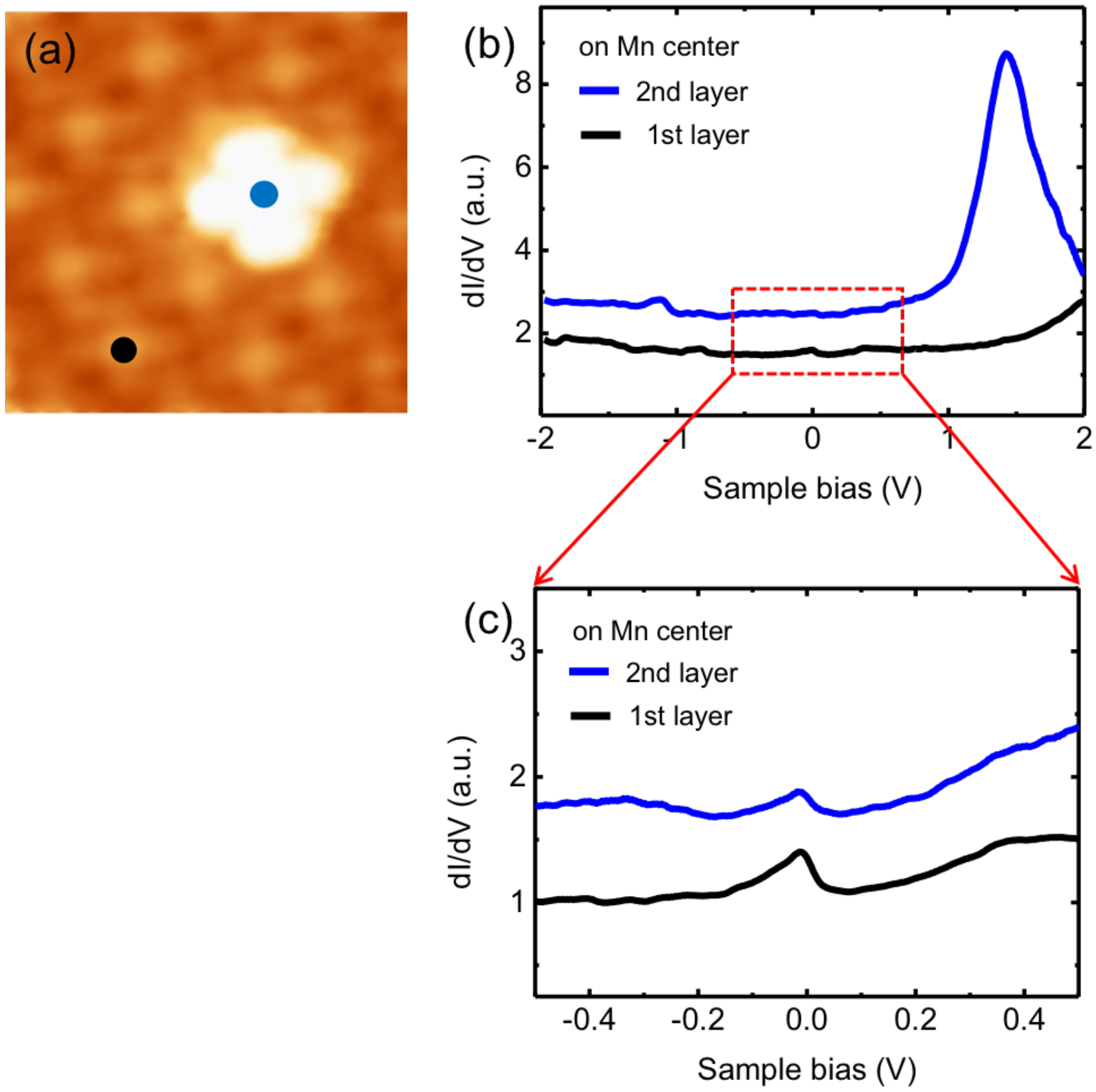

**Figure S11**. (a) STM image (6 nm x 6 nm) of a second layer MnPc molecule with *shifted* configuration (– 0.5 V, 10 pA). (b) dI/dV spectra measured on the Mn center of the first- and second-layer MnPc molecules, respectively (set point: −1.0 V, 50 pA). (c) Magnification of the spectrum in (b) close to the Fermi level (set point: −1.0 V, 100 pA). The spectra are shifted for clarity.